\documentclass[compsoc,conference,a4paper,10pt,times]{IEEEtran}
\IEEEoverridecommandlockouts
\usepackage{cite}
\usepackage{amsmath,amssymb,amsfonts}
\usepackage{algpseudocode}   
\usepackage{graphicx}
\usepackage{bm}  
\usepackage{textcomp}
\usepackage{bmpsize}
\usepackage[table,dvipsnames]{xcolor}
\usepackage{lipsum}
\usepackage{xspace}
\usepackage{multirow}
\usepackage{subcaption}
\usepackage{tikz} 
\usepackage{booktabs}
\usepackage{array} 
\usepackage{algorithm} 
\usepackage{makecell}
\usepackage{mdframed}
\usepackage{tcolorbox}

\usepackage{enumitem}
\usepackage{url}

\usepackage{pifont}
\newcommand{\cmark}{\textcolor{green!55!black}{\ding{51}}} % 
\newcommand{\xmark}{\textcolor{red!70!black}{\ding{55}}}   % 

\usepackage{amsthm}

\usepackage[nolist]{acronym}
\usepackage[colorlinks=true,urlcolor=black]{hyperref}
\def\BibTeX{{\rm B\kern-.05em{\sc i\kern-.025em b}\kern-.08em
    T\kern-.1667em\lower.7ex\hbox{E}\kern-.125emX}}

\begin{acronym}
    \acro{ML}{Machine Learning}
    \acro{ML-NIDS}[ML-NIDS]{Machine Learning-based Network Intrusion Detection Systems}
    \acro{A-NIDS}[A-NIDS]{Anomaly-based ML-NIDS}
    \acro{mlp}[MLP]{Multilayer Perceptron}
    \acro{NER}{Named Entity Recognition}
    \acro{dnn}[DNN]{Deep Neural Network}
    \acro{STD}{Standard Deviation}
    \acro{AV}{Anti-Virus}
    \acro{API}{Application Programming Interface}
    \acro{TPR}{True Positive Rate}
    \acro{FPR}{False Positive Rate}
    \acro{KG}{Knowledge Graph}
    \acro{LLM}{Large Language Model}
    \acro{CIoT-23}{CICIoT2023 Dataset}
    \acro{CIDS-17}{CIC-IDS2017 Dataset}
\end{acronym}
\begin{document}

\DeclareRobustCommand{\fullcirclednum}[1]{%
  \tikz[baseline=(char.base)]{
    \node[shape=circle,draw,inner sep=0.5pt,fill=black,text=white] (char) {\sffamily\footnotesize #1};
  }\xspace
}

\def\Snospace~{\S{}}
\renewcommand*{\sectionautorefname}{\Snospace}
\renewcommand*{\subsectionautorefname}{\Snospace}
\renewcommand*{\subsubsectionautorefname}{\Snospace}
\renewcommand*{\figureautorefname}{Figure}

\newcommand{\paragraphit}[1]{{\textit{#1.} }}
\renewcommand{\paragraph}[1]{{\vskip 4pt \noindent\textbf{#1.} }}

\newcommand{\projectTitle}{\textsc{KnowML}\xspace}

\newcommand{\KGFE}{KAFS\xspace}

\newcommand{\KITML}{EoA\xspace}

\newcommand{\RNum}[1]{\uppercase\expandafter{\romannumeral #1\relax}}

\title{\projectTitle: Improving Generalization of ML-NIDS\\with Attack Knowledge Graphs}

\author{
    \IEEEauthorblockN{Xin Fan Guo$^{\dagger  \ddagger}$, Xinran Zheng$^{\ddagger}$, Albert Merono Penuela$^{\star}$,\\Lorenzo Cavallaro$^{\ddagger}$, Sergio Maffeis$^{\dagger}$, Fabio Pierazzi$^{\ddagger}$}
    \IEEEauthorblockA{$^{\dagger}$Imperial College London, 
    $^{\star}$King's College London,
    $^{\ddagger}$University College London
    }
}

\maketitle

\begin{abstract}
Anomaly-based ML-NIDS (A-NIDS) model normal network behavior from benign data and classify deviations from this baseline as anomalies, theoretically enabling the detection of evolving attack variants without labeled attack data. The ability of A-NIDS to generalize critically depends on the quality of the feature space representing network behavior. However, the requirement for feature spaces that encode attack-relevant semantics has received little attention and remains poorly understood. As a consequence, these systems still struggle to meet practical operational constraints (low false positive rates without compromising detection performance and generalization to attack variants). We identify two limitations in the current feature spaces. First, \textit{Out-of-Dimension Blindness}, where features do not capture essential attack mechanism properties. Second, \textit{Attack Strategy Aggregation Failure}, where features cannot encode composite attack behaviors. Moreover, we demonstrate that two SotA data-driven generalization frameworks (based on incremental and contrastive learning) cannot compensate for these feature-level shortcomings. 

To bridge this gap, we present \projectTitle, a framework that encodes attack domain knowledge directly into the feature space. For each attack family, our method employs LLMs to construct a corresponding Knowledge Graph (KG) from attack implementations. Symbolic reasoning is then applied over the KG to enumerate potential attack strategies and their compositions. The resulting \textit{Knowledge-Augmented Feature Space} enables effective generalization even when trained exclusively on benign traffic, a capability beyond current approaches. Systematic empirical evaluations show that \projectTitle achieves up to 99\% detection rates while maintaining false positive rates at or below 0.0137\%, substantially outperforming contemporary feature-based baselines across diverse attack variants.

\end{abstract}


\section{Introduction}

Among \ac{ML-NIDS}, Anomaly-based ones (A-NIDS) have been widely studied for their potential to detect both evolving and previously-unseen attack variants without requiring extensive labeling to improve generalization~\cite{apruzzese2023role, zhang2024aoc, zhao2024trident}. Recent studies report strong performance on benchmark datasets, demonstrating progress in detection accuracy and generalization~\cite{zhang2024aoc, zhao2024trident, soltani2023adaptable, sarhan2022towards}, positioning A-NIDS as promising candidates for practical deployment. 

Practicality, however, imposes two demanding constraints. First, a low False Positive Rate (FPR) is essential, unlike domains such as image classification: typical network traffic rates (10,000+ pps~\cite{networkRate}) mean that even a 1\% FPR yields an impractical volume of false alarms~\cite{axelsson2000base, sommer2010outside, arp2022and}. Second, A-NIDS must effectively detect both seen and unseen attack variants. Yet contemporary work often operates under permissive FPR thresholds (e.g., 5\% in~\cite{AOC-IDS}) and evaluations are typically conducted on homogeneous datasets~\cite{flood2024bad}, which do not fully capture the diversity of attack strategies encountered in practice~\cite{nougnanke2025dataset}.

In this paper, we start by investigating whether current A-NIDS can satisfy the aforementioned operational constraints of low FPRs while detecting sophisticated attack variants. When we discuss A-NIDS, we focus on two building blocks of such systems: a feature space that represents traffic characteristics, and a machine learning model that performs anomaly-based detection. Within this setting, our initial empirical analysis indicates that the choice of feature space has a stronger influence on detection effectiveness than the choice of model architecture. This observation motivates a deeper examination of why existing feature spaces may fall short. 

We hypothesize that the underlying cause is a fundamental \textit{knowledge gap} that manifests in two forms, \emph{Out-of-Dimension Blindness}, where features directly related to attack mechanisms are missing, and \emph{Attack Strategy Aggregation Failure}, where features are not aggregated in a way that captures composite attack strategies. These gaps arise because current systems are built on incomplete and heuristic assumptions about attacker strategies. Unlike conventional ML applications such as image classification, where models operate on complete input data, that is, all pixels in an image (distorted or not) and learn representations from this entire space~\cite{rifai2011contractive, bengio2006greedy, bengio_survey}, intrusion detection operates on network traffic that contains a large and heterogeneous set of potential signals. Packet headers, timing patterns, connection states, and behavioral sequences across protocol layers are all plausible candidates for monitoring. In practice, human engineers must explicitly select which subset of these candidates to measure, and this design choice defines the feature space on which learning occurs. When the selected feature space lacks sufficient discriminative information, we observe two failure modes in our preliminary analysis (\autoref{sec:issues}). First, models require higher FPRs to reach reported detection rates, which conflicts with operational requirements for low FPRs (\autoref{subsec:fpr_inflation}). Second, attacks that primarily manifest in unmonitored features of the traffic may appear benign in the monitored features and thus evade detection, a behavior reminiscent of mimicry style attacks~\cite{wagner2002mimicry} (\autoref{subsec:fpr_inflation}). 

To close this gap, a natural idea is to enumerate a rich set of features and learn representations over this enlarged space. In NIDS, however, this is impractical for two reasons. First, real-time processing at high-packet rates hinders monitoring every potential signal; unlike malware detection, where feature spaces with millions of dimensions are feasible for static analysis~\cite{arp2014drebin}, a NIDS must balance attack-surface coverage with compactness for efficient processing. Second, data-driven feature engineering requires comprehensive datasets spanning diverse attack behaviors~\cite{li2020building, amiri2011mutual, gharaee2016new, feature_selection_using_filters}, which remain scarce in network security~\cite{flood2024bad}, causing learned representations to generalize poorly to unseen variants.

We therefore seek a feature space that is both discriminative enough to detect attack variants, and compact enough for efficient  monitoring. To achieve this, we require a proxy for approximating the space of possible attack strategies to serve as a foundation for feature design. To this end, we turn to open-source attack implementations as a concrete, analyzable record of attacker strategies and introduce \projectTitle as a framework for systematic analysis of attack implementations to extract knowledge-augmented features. 
Given an attack family name (e.g., TCP DoS), \projectTitle employs LLMs to construct a Knowledge Graph (KG) from code repositories. For example, the ``TCP DoS'' family encompasses any manipulation of TCP to achieve denial of service, where manipulating a specific field (e.g., window-size value~\cite{CVE202432984}) constitutes a variant. \projectTitle then applies  reasoning to analyze attack strategies and their combinations, deriving a \textit{Knowledge-Augmented Feature Space} (\KGFE) grounded in attack semantics.

We validate our hypothesis and the utility of \projectTitle through three complementary evaluations. First, we compare \KGFE with three widely used feature spaces across two anomaly detection models on diverse network datasets, isolating the impact of feature design from model architecture and hyperparameter choices (to avoid the known ``benchmark lottery problem''~\cite{dehghani2021benchmark}). Second, we assess whether advanced data-driven generalization techniques can substitute for a knowledge-guided feature space by comparing \projectTitle with state-of-the-art incremental and contrastive learning frameworks. Third, we conduct ablation and efficiency analyses to quantify individual component contributions and demonstrate that \KGFE remains computationally feasible while delivering the strongest overall generalization.

\noindent In this paper, we make the following contributions: 
\begin{itemize}%[leftmargin=*]
        \item We systematize the weaknesses of A-NIDS approaches, identifying two knowledge gaps arising from reliance on non-discriminative features and from an incomplete understanding of attacker strategies (\autoref{sec:issues}).
        \item We introduce \projectTitle (\autoref{sec:overview}), a framework that constructs a Knowledge Graph of attack strategies from code repositories using LLMs (\autoref{subsec:kg_construction}) and derives a \textit{Knowledge-Augmented Feature Space} grounded in attack semantics.
        \item We demonstrate through empirical evaluation that identified \textit{knowledge gaps} cannot be bridged through machine learning techniques, and propose \KGFE\ that achieves up to 99\% F1-score while maintaining FPR at or below 0.0137\% on both IoT and enterprise networks, substantially outperforming baseline approaches (\autoref{subsec:generalization}).
        \item  We release \projectTitle's 
        implementation and constructed KGs, and \KGFE\ across CICIoT2023, CIC-IDS2017.\footnote{https://github.com/s2labres/KnowML} To address the scarcity of recent attack variants in existing benchmarks, we collect, annotate, and release 9 attack variants, including exploits of recently disclosed vulnerabilities absent from standard datasets~(\autoref{sec:eval}).
\end{itemize}

\section{Motivation}\label{sec:issues}

In this section, we motivate our work by examining how current A-NIDS behave under two practical constraints, namely maintaining low false positive rates and detecting sophisticated attack variants. We show that detection performance degrades substantially when these constraints are enforced. 

\paragraph{Anomaly Detection}
Given benign traffic data $\boldsymbol{X}$ sampled from the benign distribution $\mathcal{D}$, unsupervised anomaly detection models learn normal behavior representations and assign anomaly scores to new samples. A threshold $\varphi$, determined from validation data, flags samples as anomalies when scores exceed $\varphi$ (reconstruction-based) or fall below it (density-based). An effective A-NIDS must maintain high TPR under stringent thresholds ensuring low FPR. Formally, we define ``reducing FPR'' as adjusting $\varphi$ on validation data such that FPR $\leq\varphi$~\cite{nids_thresholding}.

\paragraph{Feature-Space Baselines}
To assess how the choice of feature space affects detection performance, we compare against three widely used methods that represent distinct paradigms in the field. The first baseline is the Kitsune feature space (KIT)\cite{mirsky2018kitsune}, which provides a manually-defined set of 100 features designed from partial attack knowledge, such as using jitter in IP camera streams to detect man-in-the-middle attacks. This baseline represents the paradigm of expert-driven feature engineering based on domain-specific attack characteristics. The second baseline is the Standardized feature space (SFS)\cite{sarhan2022towards}, which derives features from CISCO NetFlow standards~\cite{cisco_ios_netflow} and extracts them using Argus~\cite{argus_ml2025}. This baseline is representative because it follows industry-standard network flow specifications and is commonly used to extract feature spaces for widely-adopted benchmark datasets such as UNSW-NB15~\cite{unsw_dataset}. The third baseline is CICFlowMeter Features (CIC)~\cite{sharafaldin2018toward} in its corrected version~\cite{9474286}, which serves as the default feature space for prominent benchmark datasets including CIC-IDS2017 and CSE-CIC-IDS2018~\cite{sharafaldin2018toward}.

\subsection{The FPR–TPR Trade off}\label{subsec:fpr_inflation}
In operational deployments, \ac{A-NIDS} must sustain very low false positive rates to avoid causing alert fatigue. Following prior work~\cite{burgio2020reduction, mirsky2018kitsune, willems2023data, shafieian2023multi}, we regard an FPR of at most 0.1\% as operationally practical.

In this section we examine how enforcing low FPR affects True Positive Rate (TPR) for \ac{A-NIDS}. We focus on two representative anomaly detectors, Gaussian Mixture Models (GMM)~\cite{an2022ensemble} and Ensembles of Autoencoders (EoA)~\cite{mirsky2018kitsune}, which we refer to collectively as baselines (for additional details on the experimental setup and results, see~\autoref{sec:eval}). Our objective is to systematically assess the respective roles of the two core components of an \ac{A-NIDS}, namely the machine learning model and its feature space.
We proceed in two stages. In the first stage, we fix the feature space and vary attack difficulty. Using KIT, we evaluate the baselines on two progressively more challenging attacks whose key behaviors are not explicitly represented in the monitored features. The first is a volumetric attack whose throughput increases during the attack (Observation 1). The second is a non-volumetric variant collected specifically for this study (Observation 2), both variants aim to achieve the same outcome--DoS. In the second stage, we fix the ML-model  and vary the feature space to assess the contribution of feature design. We compare three widely used feature spaces (CIC, SFS, KIT) (Observation 3). The complete set of results for all feature spaces is presented in \autoref{sec:eval}.

\begin{table}[!t]
\small
\centering
\caption{\textbf{Detection Performance (\%) of Two Baseline Anomaly Detectors on \ac{CIDS-17}}. $\text{FPR}_{tr}$ denotes the target \ac{FPR} used for threshold tuning, and $\text{FPR}_{te}$ the rate measured on the test set.}
\label{tab:high_fpr}
\setlength{\tabcolsep}{2.5pt}
\begin{tabular}{|c|c|c|c|c|c|} 
\hline
\textbf{$\text{FPR}_{tr}$}    & \begin{tabular}[c]{@{}c@{}}\textbf{Anomaly}\\\textbf{Model}\end{tabular} & \textbf{Metric} & \begin{tabular}[c]{@{}c@{}}\textbf{DoS}\\\textbf{GoldenEye}\end{tabular} & \begin{tabular}[c]{@{}c@{}}\textbf{DoS}\\\textbf{Hulk}\end{tabular} & \begin{tabular}[c]{@{}c@{}}\textbf{Benign }\\\textbf{$\text{FPR}_{te}$}\end{tabular}  \\ 
\hline
\multirow{4}{*}{$\leq10.0$}     & \multirow{2}{*}{GMM~}                                                    & TPR          & 93.38                                                                   & 93.72                                                               & \multirow{2}{*}{11.02}                                                                      \\ 
\cline{3-5}
                              &                                                                          & F1-score        & 92.71                                                                    & 96.26                                                               &                                                                                            \\ 
\cline{2-6}
                              & \multirow{2}{*}{EoA}                                                 & TPR          & 72.62                                                                    & 57.15                                                               & \multirow{2}{*}{9.98}                                                                      \\ 
\cline{3-5}
                              &                                                                          & F1-score        & 78.99                                                                    & 72.16                                                               &                                                                                            \\ 
\hline
\multirow{4}{*}{$\leq 0.1$} & \multirow{2}{*}{GMM~}                                                    & TPR          & 17.59                                                                   & 10.15                                                                 & \multirow{2}{*}{0.10}                                                                      \\ 
\cline{3-5}
                              &                                                                          & F1-score        & 29.90                                                                    & 18.43                                                               &                                                                                            \\ 
\cline{2-6}
                              & \multirow{2}{*}{EoA}                                                 & TPR          & 63.87                                                                    & 38.94                                                                & \multirow{2}{*}{0.10}                                                                      \\ 
\cline{3-5}
                              &                                                                          & F1-score        & 77.92                                                                    & 56.04                                                               &                                                                                            \\
\hline
\end{tabular}
\end{table}

\begin{figure}[t!]
  \centering
  \begin{subfigure}[t]{\linewidth}
    \centering
    \includegraphics[width=\linewidth]{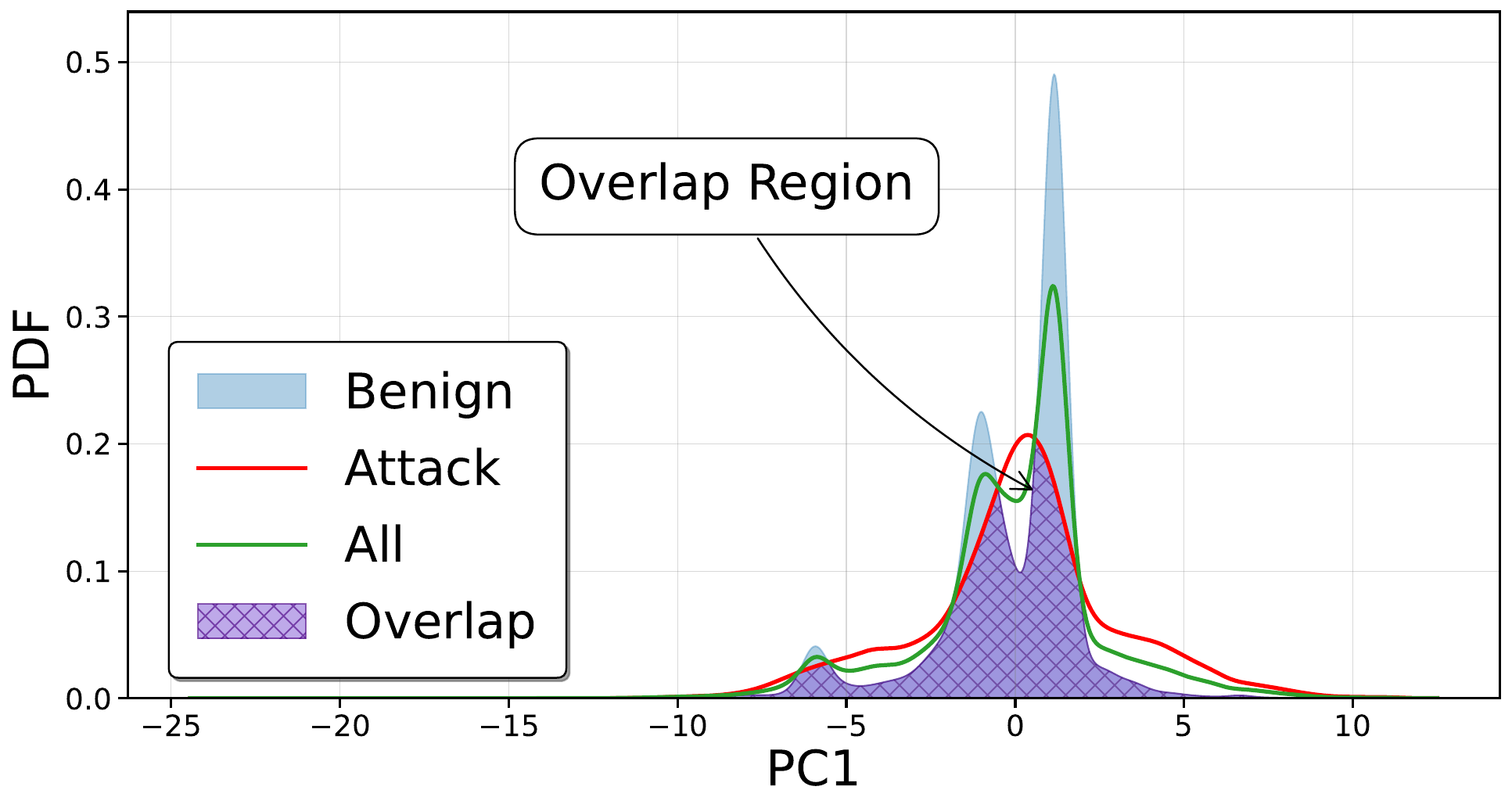}
    \caption{Partial overlap: DoS-Hulk vs. benign traffic}
    \label{fig:fcta}
  \end{subfigure}
  \begin{subfigure}[t]{\linewidth}
    \centering
    \includegraphics[width=\linewidth]{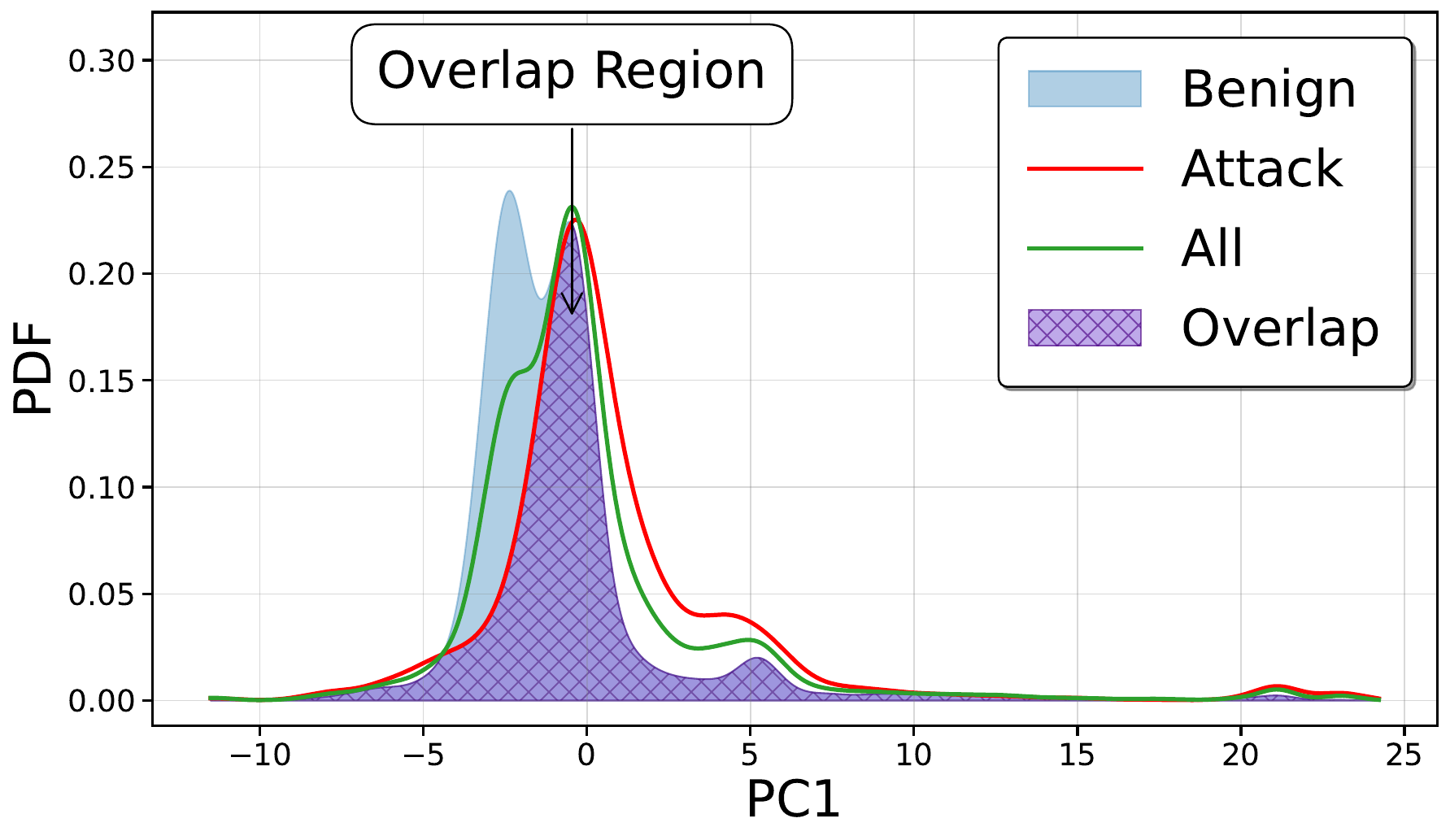}
    \caption{Substantial overlap: Nkiller2 vs. benign traffic}
    \label{fig:lengtha}
  \end{subfigure}
  \caption{\textbf{Attack-Benign Distributional Overlap in KIT feature space.} Probability density distributions projected onto the first principal component (PC1). (a) DoS-Hulk exhibits partial overlap with benign traffic, forcing a TPR-FPR trade-off. (b)  Nkiller2 exhibit near-complete overlap, rendering detection infeasible at low FPR.}
  \label{fig:roverlap}
\end{figure}

\noindent\textbf{Observation 1: TPR degrades when attack traffic partially overlaps with benign under constrained FPR.} 
When enforcing operationally realistic FPR constraints ($\leq$0.1\%), testing on attack variants that partially overlap with benign traffic (\autoref{fig:fcta}) leads to substantial TPR degradation. This effect is not always visible in prior evaluations, which often adopt higher FPR thresholds, for example, 5\% in~\cite{AOC-IDS}, or emphasize other metrics such as Precision (for example, Precision up to 97.25\% in~\cite{TridentCode}) without explicitly constraining FPR~\cite{zhai2018random, ferrag2020deep, racherla2024deep, kim2022robust}. For illustration, consider GMM~\cite{an2022ensemble} and \KITML~\cite{mirsky2018kitsune}, both trained on benign CIC-IDS2017 traffic~\cite{sharafaldin2018toward} and evaluated on DoS-GoldenEye and DoS-Hulk (HTTP DoS variants). Under an FPR constraint of at most 0.1\%, GMM TPR on DoS-GoldenEye decreases from 93.38\% to 17.59\%, and \KITML\ TPR on DoS-Hulk decreases from 57.15\% to 38.94\% (\autoref{tab:high_fpr}). Similar patterns under strict FPR constraints are reported in~\cite{hashemi2020enhancing}, suggesting that maintaining high TPR at low FPR remains a systemic challenge for current \ac{A-NIDS} approaches.

\noindent\textbf{Observation 2: TPR collapses under substantial attack–benign overlap and constrained FPR.}
Performance degradation becomes more pronounced when attack variants closely resemble benign traffic, particularly for attacks that do not cause significant changes in network throughput. The Nkiller2 DoS attack~\cite{CVE202432984} illustrates this challenge. Nkiller2 manipulates the TCP window size while preserving packet lengths, inter-arrival times, and throughput, features that are central to many existing detection methods~\cite{mirsky2018kitsune, 8984222, hypervision, whisper}. By maintaining throughput patterns that are statistically similar to benign traffic, Nkiller2 can evade detectors that rely primarily on volume-based anomalies. This effect is visible when projecting Kitsune’s feature space onto its most discriminative principal component (PC1), where Nkiller2 traffic coincides with benign traffic (\autoref{fig:lengtha}). Under low FPR constraints, this overlap leads to detection failure. On CICIoT2023, Kitsune attains 99.88\% F1 score on conventional TCP DoS attacks yet yields 0\% F1 score on Nkiller2. This discrepancy highlights that current \ac{A-NIDS} may struggle to generalize to operationally realistic variants whose behavior is not adequately captured by the monitored features.

\begin{figure}[t!]
    \centering
    \includegraphics[width=0.9\linewidth]{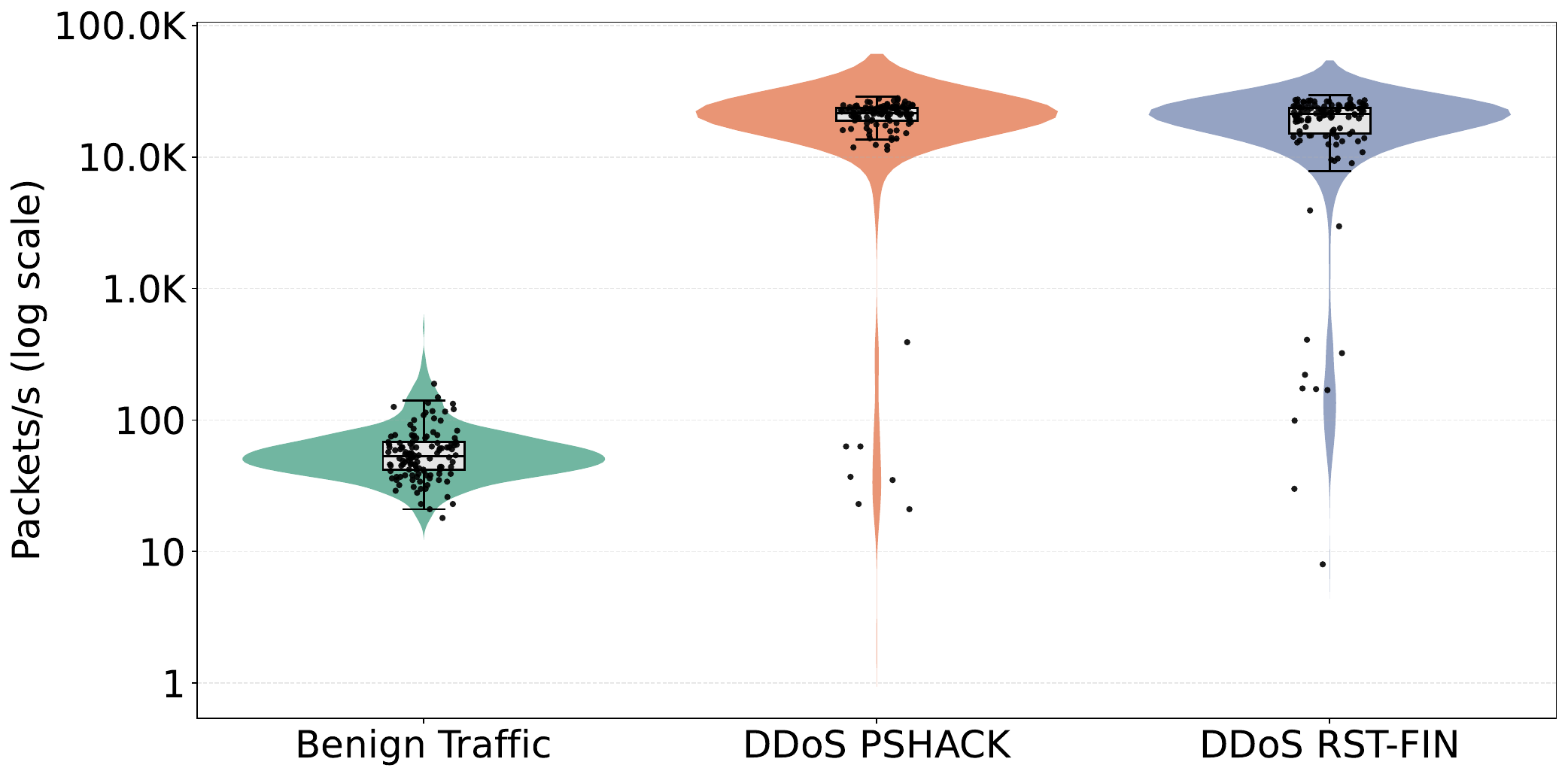}
    \caption{\textbf{Statistical Separation in Packet Rate Distributions.} DDoS attacks (DDoS-SYN/ACK Flood and DDoS-RST/FIN Flood) on CICIoT2023 exhibit substantially elevated and concentrated packet rates compared to benign traffic, representing statistically obvious attack signals that should enable straightforward detection.}
    \label{fig:obsv3}
\end{figure}

\noindent\textbf{Observation 3: Inconsistent TPR across feature spaces for attacks that are statistically-distinct from benign.}
Under the same FPR constraints, baseline models can exhibit inconsistent performance even on attacks that are statistically distinct from benign traffic. We evaluate \KITML\ with hyperparameter optimization on three feature spaces for CICIoT2023 DDoS attacks. As shown in \autoref{fig:obsv3}, these attacks display markedly elevated packet rates with clear distributional separation from benign traffic. Although all three feature spaces include packet/s features, detection performance differs substantially (\autoref{tab:obsv3}). KIT achieves 98.31\% and 99.50\% F1-scores on DDoS PSHACK and DDoS RSTFIN, respectively, whereas the SFS~\cite{sarhan2022towards} and CIC~\cite{sharafaldin2018toward} obtain 0\% F1-score on both attacks. Upon inspection of the evaluated feature spaces, it reveals that KIT aggregates packet size and rate at different endpoint levels, suggesting that the way traffic statistics are aggregated, rather than their mere presence, plays a critical role in effective detection for \ac{A-NIDS}.

\begin{table}[t!]

   \caption{\textbf{Detection Inconsistency Across different Feature Spaces.} F1-scores (\%) on CICIoT2023 DDoS attacks under FPR ($\leq0.1\%$). Applying identical model architecture (\KITML) with optimized hyperparameters. Despite all feature spaces containing packet rate features and attacks exhibiting obvious statistical separation, different feature spaces exhibit markedly different detection capabilities.}\label{tab:obsv3}

    \centering
    \begin{tabular}{l|c|c|c}
    \hline
    Attack & SFS & CIC & KIT \\
    \hline
    DDoS-PSHACK (CIoT-23)  & 0.00 & 0.78 & 98.31 \\
    DDoS-RSTFIN (CIoT-23)  & 0.00 & 0.00 & 99.50 \\
    \hline
    \end{tabular}

\end{table}

\newcommand\kgap[1]{Gap~\RNum{#1}} %SM suggestion

\subsection{Identified Knowledge Gaps}\label{subsec:knowledge_gap} 
The three observations indicate that current \ac{A-NIDS} can be limited by \textit{inadequate feature-space design}. In particular, available features may lack attack-relevant semantics or appropriate aggregation granularity to reliably separate benign from malicious traffic under practical FPR constraints. These limitations can be viewed as arising from the following two knowledge gaps.

\paragraph{\textbf{\kgap{1}: Out-of-Dimension Blindness}}\label{subsec:out_of_dim} The first gap appears when attacks exploit dimensions that are not represented in the feature space.  Human-designed feature spaces are constrained by researchers' knowledge of plausible attack strategies and specific detection tasks. When discriminative dimensions are absent, attack traffic can become statistically indistinguishable from benign traffic, which is reflected in Observations 1 and 2.  The Nkiller2 attack exemplifies this behavior. Kitsune, for instance, monitors jitter and packet size but does not include TCP window-related features, leading to near-complete overlap between Nkiller2 and benign traffic in the monitored space. The feature space illustrated in \autoref{fig:roverlap} similarly lacks HTTP protocol semantics, which makes HTTP-based attacks difficult to distinguish from benign traffic when throughput does not differ substantially (Observation 1). Under such conditions, achieving high TPR at low \ac{FPR} becomes infeasible: the substantial overlap between attack and benign distributions allows for no decision boundary that improves TPR without incurring excessive false alarms.

\begin{figure}[!t]
    \centering
    \includegraphics[width=0.9\linewidth]{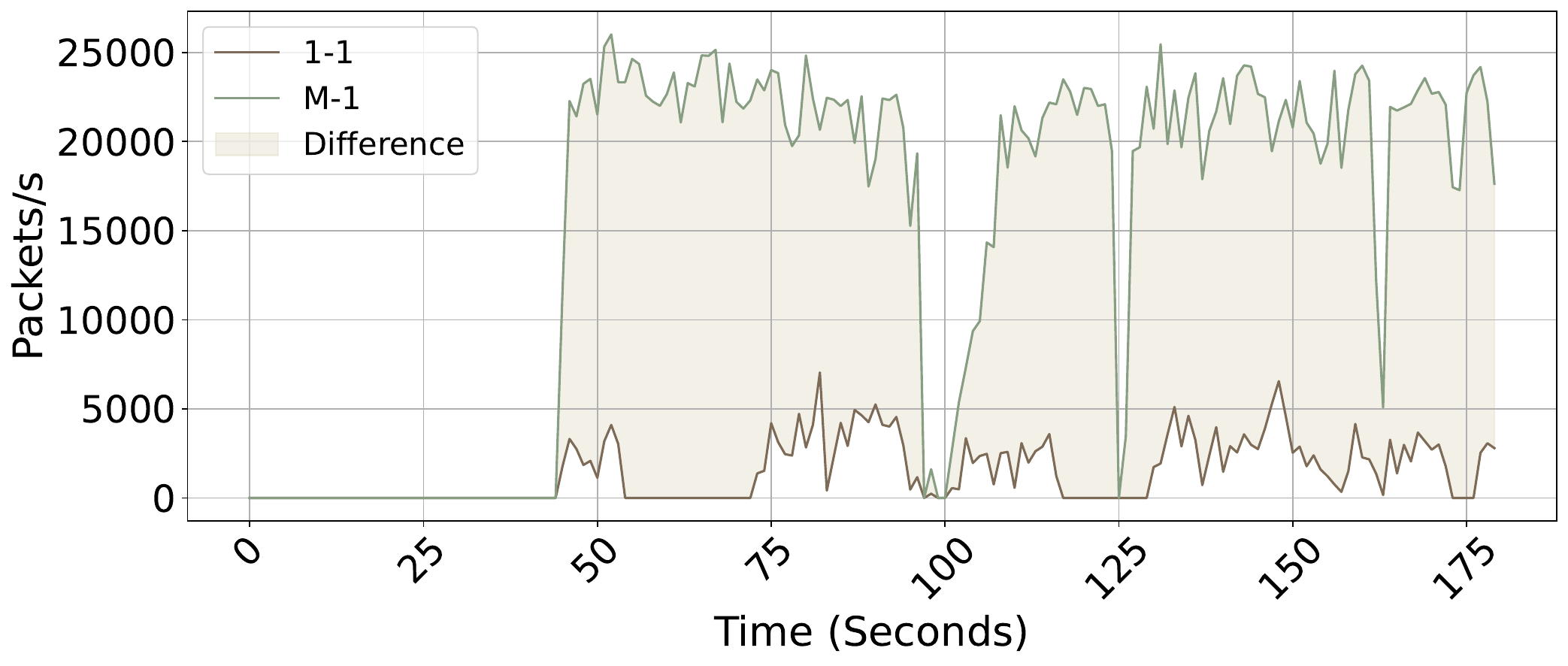}
    \caption{
        \textbf{Detection Failure Under Per-Flow Analysis.} 
        RST-FIN packets/s during distributed DoS. Individual flow monitoring (1:1, brown) shows benign-appearing traffic below detection thresholds. Aggregating multiple sources to single victim (M:1, green) exposes attack bursts (shaded regions) invisible to per-flow feature space.
    }
    \label{fig:aggregation}
\end{figure}

\paragraph{\textbf{\kgap{2}. Attack Strategy Aggregation Failure}}\label{subsec:strategy_aggregation} The second critical gap arises when attackers employ combined strategies where individual attack components remain invisible when monitored individually. A canonical example is a multi source distributed DoS attack. Many contemporary \ac{A-NIDS} are trained on feature spaces that provide per-flow statistics and analyze each connection in isolation. When monitoring traffic to a single victim ($N = 1$), short-lived TCP connections from multiple sources often appear benign on a per-flow basis and remain below the anomaly threshold $\varphi$. In contrast, aggregating connections across sources ($M$ attackers to $N = 1$ victim) reveals clear bursts, as illustrated in \autoref{fig:aggregation}, where $M\!:\!1$ temporal aggregation exposes patterns that are invisible in individual $1\!:\!1$ flows. Observation 3 demonstrates this effect. Even when attacks exhibit clear statistical signals, such as elevated packet rates in \autoref{fig:obsv3}, detection success depends on how features aggregate traffic. For example, an $M\!-\!N$ feature that sums corresponding $1\!-\!N$ flows sharing the same destination can make a distributed attack salient, whereas purely per-flow features may not. In our experiments, the same packet rate signal yields a 99.50\% F1 score under one aggregation scheme (KIT) but 0\% under others (SFS, CIC), despite identical underlying traffic and model architecture (\autoref{tab:obsv3}). Although distributed attacks are well studied~\cite{prajapati2014literature, shurman2020and, akgun2022new}, this knowledge is not always reflected in feature space design, which can hinder the detection of combined attack strategies. These limitations do not invalidate existing feature spaces but indicate that they are insufficient on their own under realistic operational constraints.

\newcommand\ttype[1]{{\small \texttt{#1}}}

\section{Our Approach: Leveraging Attack Knowledge for Systematic Generalization}
\label{sec:our_approach}
Current A-NIDS fall short because human-designed feature spaces cannot capture the diverse attack surfaces adversaries exploit (\kgap{1}) nor represent  the combined strategies attackers employ (\kgap{2}). We propose \projectTitle, which addresses both limitations by mining known attack implementations to systematically enumerate attack dimensions and their combinations through a Knowledge Graph (KG), and 
 translating these insights into Knowledge-Augmented Features (\KGFE)\ that encode both atomic attack tactics and their composite strategies. Unlike human-designed features that rely on researcher intuition about plausible attacks, our approach grounds feature space construction in \emph{actual attacker behavior} extracted from relevant open-source attack implementations, ensuring coverage of dimensions and aggregations that baselines miss.

We construct this KG from publicly available attack implementations. This approach is practical because adversaries demonstrably reuse attack mechanisms across time. Analysis of NIST vulnerability databases reveals recurring patterns; e.g., the previously mentioned Nkiller2 DoS appeared in 2008, 2009, and 
2024~\cite{CVE20084609,CVE-2009-1926,CVE202432984}; Slowloris connection throttling was reported in 2007 and resurfaced in 2025~\cite{NVD_CVE_2007_6750,NVD_CVE_2025_32472}; a 2025 HTTP denial-of-service attack combined a 2024 header exploitation with  request smuggling from 2020~\cite{CVE-2025-31650,CVE-2024-35296,CVE-2020-35863}. While this analysis does not claim comprehensive coverage of all attack variants, these examples illustrate that past attack knowledge provides actionable primitives for anticipating emergent threats through recombination.

To extract these behavioral primitives, we analyze launch parameters from attack 
implementations' \ttype{README.md} documentation. Launch parameters are command-line 
arguments controlling attack execution, typically documented as:

{\small
\begin{verbatim}
$ python script.py --keep-alive 
                   --syn_flag
  --keep-alive: Maintain persistent \
                connections
  --syn_flag: Send SYN packets with \
              specified flags
\end{verbatim}
}

\begin{figure}[t]
  \centering
  \begin{subfigure}[t]{0.5\linewidth}
    \centering
    \includegraphics[width=\linewidth]{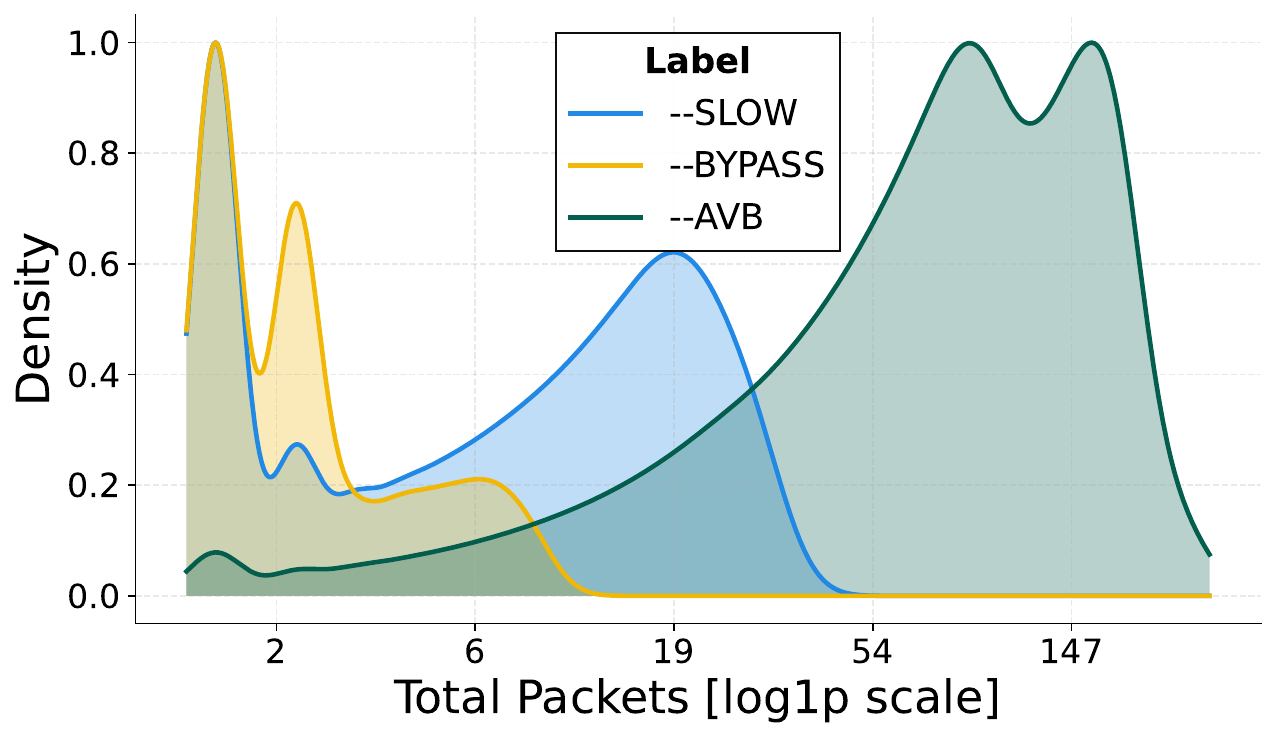} % e.g., figs/fct_dist.pdf
    \caption{Total Packet Count}
    \label{fig:fct}
  \end{subfigure}\hfill
  \begin{subfigure}[t]{0.5\linewidth}
    \centering
    \includegraphics[width=\linewidth]{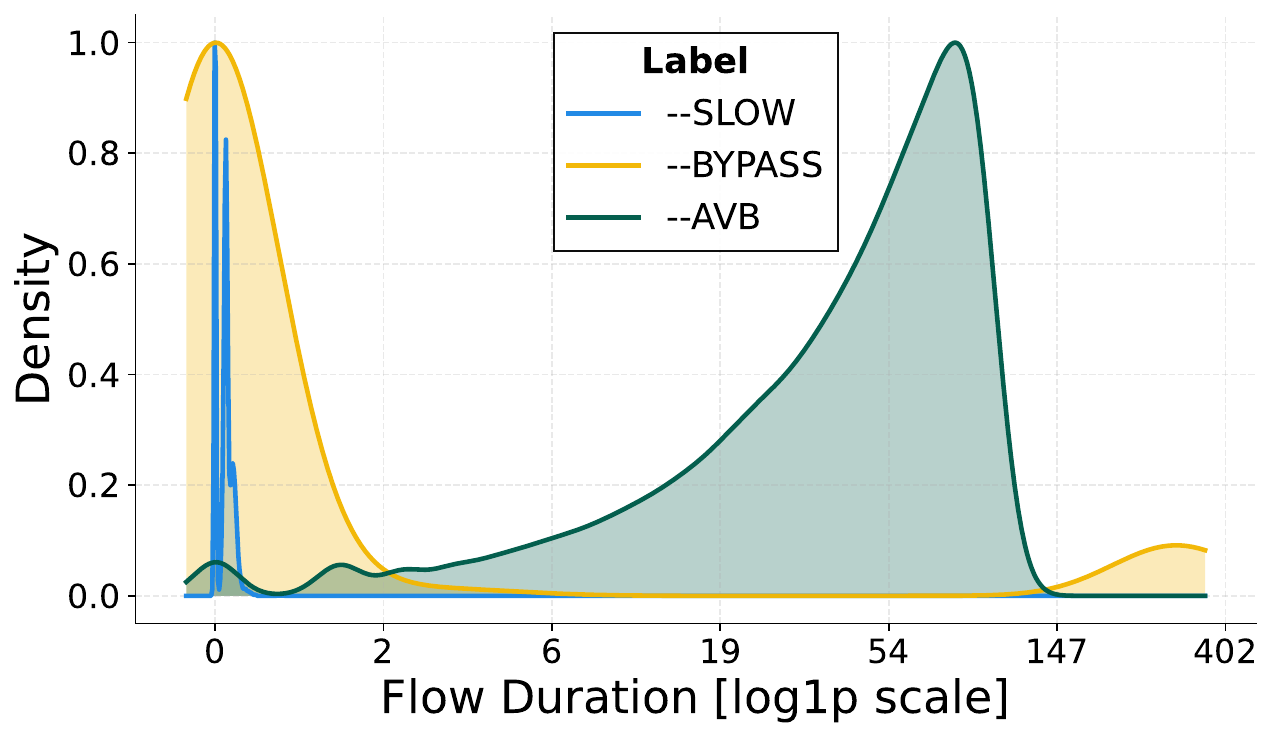} % e.g., figs/flow_len_dist.pdf
    \caption{Flow Duration}
    \label{fig:length}
  \end{subfigure}
      \caption{\textbf{Launch parameter variations in HTTP DoS attacks~\cite{fiberfox} produce distinct traffic distributions.} Different settings (SLOW,  
      BYPASS, and AVB for details see~\cite{fiberfox}) yield measurably different behaviors in (a) Packet Count and (b) Flow Duration, demonstrating that parameters encode operationally significant attack strategies.}
  \label{fig:set_param}
\end{figure}

These parameters encode behavioral attack strategies specifying how attackers vary their 
techniques. For example, launching a TCP DoS attack with \ttype{--keep-alive} 
and \ttype{--syn\_flag} prolongs resource exhaustion by preventing connection 
timeouts while flooding SYN queues (see~\autoref{subsec:kg_construction} for how we remove unrelated parameters). Launch parameters effectively represent attack 
variations, as demonstrated by their use in prior work for generating attack variants in generalization testing~\cite{concap}. As shown in 
\autoref{fig:set_param}, different parameter combinations for HTTP DoS attacks 
(SLOW, BYPASS, AVB) produce measurably distinct traffic patterns in packet count 
and flow duration distributions. This variability confirms that launch parameters 
encode operationally relevant behaviors, making them effective primitives for 
enumerating attack strategies and analyzing their compositions.

\section{\projectTitle}\label{sec:methodology}

%\paragraph{Design Goals}
Informed by the knowledge gaps and motivation in~\autoref{sec:issues} and~\autoref{sec:our_approach}, \projectTitle pursues two core objectives. First,  we construct a Knowledge Graph of attack strategies organized by attack family, as different protocols expose distinct attack surfaces through their states, header structures, and connection semantics. Second, 
we analyze how these strategies combine through parameter compositions to reveal 
attack variants emerging from combined strategies. Note that in this work we focus on automatically extracting attack strategies and analyzing their combinations. We intentionally retain a human in the loop when mapping identified strategies to measurable features, which enables the incorporation of domain-specific expertise that current LLMs do not capture in a sufficiently reliable or fully automatic manner. We discuss the implications of this design choice in \autoref{sec:discussion}.

\paragraph{Threat Model}
We focus on active network attacks that generate observable traffic patterns at 
the flow level, consistent with the operational constraints of anomaly-based NIDS. Our KG is constructed from publicly available attack implementations, ensuring reproducibility while covering widely deployed threat classes (a more detailed discussion of adversaries in this space is provided in \autoref{sec:discussion}). We assume an attacker who uses automated scripts to deploy one or more known attack strategies. We exclude three categories from scope. First, we exclude passive reconnaissance attacks 
(e.g., off-path TCP inference~\cite{feng2020off, feng2022off_usenix}), which produce no anomalous flow-level traffic. Second, we exclude content-dependent attacks requiring Deep Packet Inspection (DPI), such as SQL Injection or Command Injection, as these operate at the payload level beyond flow-based monitoring. Third, we exclude entirely novel strategies with no publicly-available implementation, as our approach is grounded in observable implementation characteristics.

\paragraph{Background: Knowledge Graphs and Symbolic Reasoning}
In contrast to statistical and data-driven methods that learn from raw data, symbolic reasoning operates on structured, human-interpretable knowledge to deduce general rules~\cite{zhang2021neural}. It models relationships between abstract concepts and integrates structural and semantic knowledge, enabling relational dependence analysis of concepts.

A primary tool for implementing symbolic reasoning is the \acf{KG}. A \ac{KG} is a directed graph composed of subject-predicate-object $(s, p, o)$ triples, where each triple defines a relationship between entities~\cite{kendall2019ontology}. We formally define our \ac{KG} as $G = (V, E)$, where nodes $V = S \cup R \cup F$ consist of strategy nodes $S = \{s_1, \ldots, s_n\}$ representing attack techniques, repository nodes $R = \{r_1, \ldots, r_m\}$ representing repository implementations, and attack family nodes $F = \{f_1, \ldots, f_k\}$ representing attack families. The primary edge set $E_{SR} \subseteq S \times R$ links strategies to their implementing repositories. After clustering semantically similar strategies into a partition 
$\Pi = \{C_1, \ldots, C_p\}$ of $S$, where each cluster $C_i \subseteq S$ contains 
related strategies (\autoref{subsec:kg_compression}), we select one representative 
strategy $s_j$ per cluster to form the atomic strategy set 
$\mathcal{A} = \{s_1, \ldots, s_p\}$. Symbolic reasoning rules 
(\autoref{subsec:symbolic_reasoning}) then derive transitive edges 
$E_{\mathrm{trans}} \subseteq \mathcal{A} \times \mathcal{A}$ capturing strategy 
co-occurrence patterns. 

\begin{figure*}[t]
    \centering
    \includegraphics[width=1\textwidth]{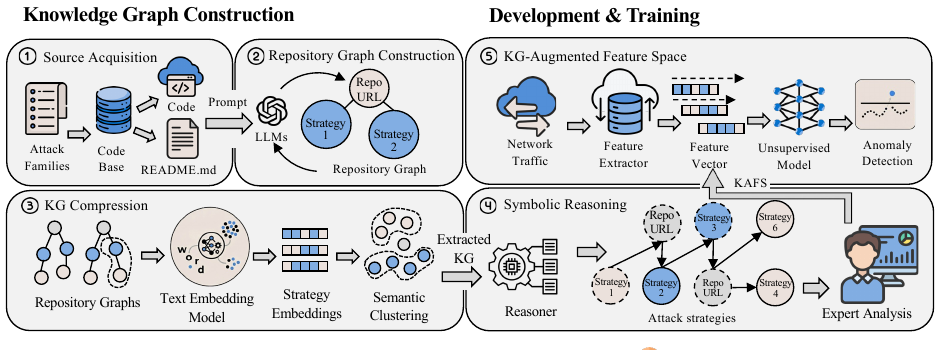} % Replace with your image filename
    \caption{\textbf{Overview of the \projectTitle Pipeline from Attack Implementation to Feature Generation.} 
    Given an attack name, the system retrieves its implementation and constructs repository graphs, which are then unified into a \ac{KG} (Steps 1–3). Symbolic reasoning is applied over the \ac{KG} to infer atomic, and composite strategies (Step 4). These results are translated into \KGFE that serve as inputs to \ac{A-NIDS} for training and evaluation (Step 5).}
    \label{fig:chrollo_pipeline}
\end{figure*}

\subsection{Overview of \projectTitle}\label{sec:overview}

\projectTitle is a framework for constructing KG of attack strategies and inferring 
Knowledge-Augmented Features (\KGFE) that directly address the generalization challenge of A-NIDS under operational constraints. The overall pipeline is shown in \autoref{fig:chrollo_pipeline} and consists of the following stages:

\begin{itemize}
\item \textbf{\ac{KG} Construction} \fullcirclednum{1} -- \fullcirclednum{3} (\autoref{subsec:kg_construction}): Open-source attack repositories are automatically parsed to build a \ac{KG} given an attack family name. This step unifies identified attack strategies and selects representative strategies to build the final KG. 

\item \textbf{Symbolic Reasoning} \fullcirclednum{4} (\autoref{subsec:symbolic_reasoning}): Inference rules are applied over the \ac{KG} to enumerate possible attack strategies and analyze their combinations through transitive relations. 
\item \textbf{Knowledge-Augmented Feature Space (\KGFE)} \fullcirclednum{5} (\autoref{subsec:kg_augmented}): Identified strategies and their combinations are translated into features through defined rules. 
\end{itemize}

\subsection{KG Construction}\label{subsec:kg_construction}
We construct a \ac{KG} by systematically extracting attacker strategies from open-source implementations. Specifically, we extract parameter values and their description from \ttype{README.md} (see~\autoref{sec:our_approach} for rationale). Note that here we intentionally avoid examining the specific values assigned to these parameters (e.g., \ttype{--keep-alive=0} vs. \ttype{--keep-alive=1}) and only capture the parameter name and its description. This abstraction avoids overfitting to narrow configurations and enables generalization to variants generated by different parameter value combinations (empirically validated in~\autoref{sec:eval}). The next section describes how we (1) acquire relevant repositories, (2) construct graphs from individual repositories, and (3) merge them into a unified \ac{KG}.

\subsubsection{\texorpdfstring{Source Acquisition \fullcirclednum{1}}{Source Acquisition (1)}}
We design a pipeline to mine open-source repositories for diverse attack implementations (see \fullcirclednum{1} in \autoref{fig:chrollo_pipeline} and details of the acquisition process are provided in~\autoref{app:source_acq}). Given an 
attack family name (e.g., ``HTTP DoS''), the pipeline queries the GitHub  API to search~\cite{github-rest-search} repository names and descriptions. To account for naming variations, we combine the family name with protocol-specific keywords such as ``HTTP Flood,'' ``HTTP GET,'' and ``HTTP Overload''. We retrieve \ttype{README.md} files from matched repositories, filtering out those with missing or empty documentation. These files serve as the primary source for constructing Repository Graphs in subsequent steps. GitHub-based acquisition was selected over alternative sources (e.g., NIST NVD, security advisories) because it provides structured, parsable documentation (\ttype{README.md} files) amenable to large-scale automated extraction. While vulnerability databases contain valuable information, they typically provide only brief descriptions of encountered vulnerabilities without the 
behavioral parameters necessary for our analysis. The implications of this design choice, including coverage limitations, are discussed in~\autoref{sec:discussion}.

\begin{figure}[t!]
    \centering
    \includegraphics[width=1\linewidth]{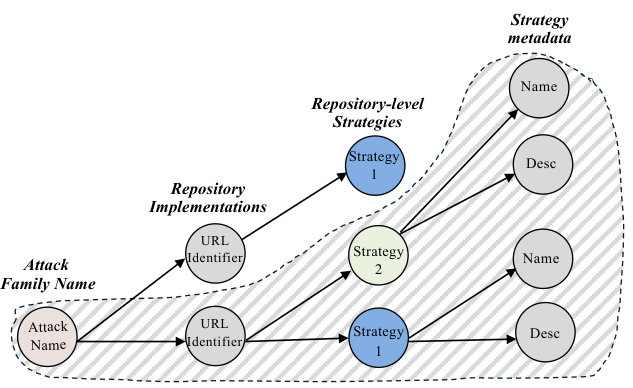}
    \caption{\textbf{Repository Graph Structure.} Each Repository Graph 
    contains: 1) a shared Attack Family Name root node (e.g., 
    ``HTTP DoS''), 2) URL Identifier nodes representing individual 
    repositories, and 3) Strategy nodes extracted from each 
    repository, where each strategy has Name and Description (`Desc')
    properties. Dashed borders delineate individual repositories. The two 
    blue-tinted Strategy 1 nodes represent semantically similar strategies 
    extracted from different repositories, which will be merged during KG 
    compression (\autoref{subsec:kg_compression}).}
    \label{fig:repo_graph}
\end{figure}

\subsubsection{\texorpdfstring{Repository Graph Construction \fullcirclednum{2}}{Repository Graph Construction (2)}}\label{subsec:repo_contruction}
In this step, we transform individual attack implementations retrieved in \fullcirclednum{1} into Repository Graphs. To achieve this, we define an ontology (\ac{KG} schema) following best practices in ontology engineering~\cite{kendall2019ontology}. The schema specifies four key classes and properties: (i) \textit{Attack Family Name} (e.g., TCP DoS), (ii) \textit{Strategy} (a specific technique employed by attackers), (iii) \textit{Description} (a textual explanation of how the strategy configures the attack), and (iv) \textit{Repository Identifier} (a URI linking each strategy to its source repository). The URI is a functional property, ensuring a one-to-one mapping between strategies and repositories. The Repository graph is illustrated in~\autoref{fig:repo_graph}. We cast the extraction problem as \ac{NER} and employ \acp{LLM} to identify entities and relations from code and documentation at scale. 
Prior work has demonstrated that \acp{LLM} are effective for a range of security tasks~\cite{aldaihan2025clouseau}, including security-oriented \ac{NER} applications~\cite{sainz2023gollie, giglou2024llms4om, 10.1145/3605943, edge2024localglobalgraphrag}. We adopt a GraphRAG-style approach~\cite{edge2024localglobalgraphrag}, enabling LLMs to extract the most relevant entities. Using the GPT-4o-mini model, our implementation achieves 90.91\% Recall on our NER benchmark (see~\autoref{app:llm_evl} for details) comparable to state-of-the-art systems especially designed for NER tasks~\cite{sainz2023gollie}. This benchmark is based on a human-annotated dataset we created for evaluation, and the high Recall demonstrates strong agreement with human annotations. This design enables large-scale, automated construction of Repository Graphs, previously infeasible with manual analysis. The subsequent paragraphs discuss how \projectTitle mitigates issues related to scalability, Recall degradation (due to increased context-window sizes), and \ac{LLM} hallucinations.

\paragraph{\textbf{Logical Consistency and Tractable Reasoning}} 
To ensure that the constructed KG remains scalable and suitable for symbolic reasoning, we defined the ontology in accordance with OWL Lite standards~\cite{w3c-owl-features-2004}, which guarantee polynomial-time reasoning under Description Logic (DL). We further validated the ontology using the HermiT reasoner~\cite{dentler2011comparison}, confirming that (i) no contradictions exist, (ii) all concepts are satisfiable, and (iii) all axioms are mutually compatible.

\paragraph{\textbf{Scalability and Recall Preservation}} 
A challenge in extracting attack strategies arises from the unpredictable length of \ttype{README.md} files and source code, which can degrade \ac{LLM} Recall as context size increases~\cite{kuratov2024search, liu2024lost}. To address this, we implemented an iterative ``gleaning'' method. After the initial extraction, the \ac{LLM} performs binary (\ttype{YES/NO}) checks for missed entities, guided by logit bias to enforce confirmation. If the model returns \ttype{YES}, an additional extraction round is triggered to extract missed entities~\cite{edge2024localglobalgraphrag}.

\paragraph{\textbf{Managing Hallucination and Noise}} LLMs are known to hallucinate, i.e., generate inaccurate or fabricated entities~\cite{huang2025survey}. To mitigate this, we employed a few-shot prompting strategy with explicit examples that cover the full spectrum of expected inputs: (i) \textit{Structured inputs}, where extraction is straightforward (e.g., option lists in code or documentation: \verb|python script.py [-p1] [-p2]| with \verb|-p1| and \verb|-p2| explained); (ii) \textit{Unstructured inputs}, where strategies are described in free-form text; (iii) \textit{Empty inputs}, where no parameters are explicitly mentioned (e.g., simply \ttype{\$ python script.py}); and (iv) \textit{Irrelevant inputs}, where search results contain unrelated repositories (e.g., an HTTP Redis pooler returned in response to an HTTP DoS query~\cite{serverless-redis-http}). In cases where the LLM extracts unrelated parameters (for example, \texttt{threads=1}), we first cluster related inputs using the procedure in~\autoref{subsec:kg_compression} and then manually discard the irrelevant clusters. This is efficient because removal is performed at the level of clusters that correspond to unique parameters. Moreover, we enforced structured output templates aligned with our ontology~\cite{openai-structured-outputs}, placing each entity into predefined fields. By constraining the \ac{LLM} to produce structured outputs and applying few-shot prompting covering diverse scenarios, \projectTitle achieves over 90\% Recall in the NER task, matching human-annotated data and reducing hallucination (see~\autoref{app:llm_evl}).

\subsubsection{\texorpdfstring{KG Compression \fullcirclednum{3}}{KG Compression (3)}}\label{subsec:kg_compression}
To eliminate redundant strategy nodes, we compress the extracted Repository Graphs by clustering similar strategies, e.g., ``Randomize Ports'' and ``Random Port Targeting''. Each strategy is embedded using its Name and Description via OpenAI’s text-embedding model~\cite{openai2024embeddings}, and clustering is performed with Hierarchical Agglomerative Clustering (HAC)~\cite{mullner2011modern}. We adopt HAC with complete linkage, as it maximizes inter-cluster separation and yields more fine-grained clusters compared to alternatives~\cite{diaz2025k}, ensuring that semantically distinct strategies are not merged. Here, we prioritize inter-cluster distance over intra-cluster distance, ensuring that the worst case produces a few repetitive clusters rather than losing unique attack strategies. For each cluster $C_j$, we select a representative strategy $S_j$ defined as the node with minimum cumulative embedding distance to all others in the same cluster:

\begin{equation}
    S_j = \arg\min \{ \sum_{i : s_i \in C_j} \left\| e_i - e_\ell \right\| \;\\:  s_\ell \in C_j \},
\label{eq:cluster}
\end{equation}
\noindent where $e_i$ is the embedding of strategy $s_i$. This representative preserves the core semantics of the cluster while eliminating redundant variants.

\subsection{\texorpdfstring{Symbolic Reasoning \fullcirclednum{4}}{Symbolic Reasoning (4)}}\label{subsec:symbolic_reasoning}
\begin{figure}[t]
    \centering
    \includegraphics[width=0.83\linewidth]{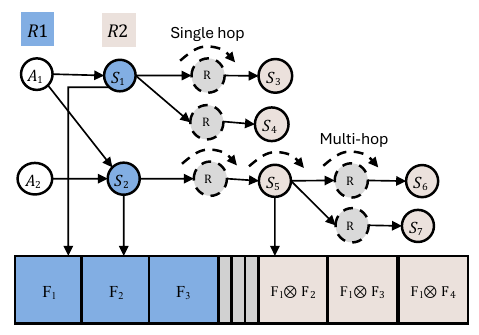}
    \caption{\textbf{Example of Symbolic Reasoning Rules (top) to derive Knowledge-Augmented Feature Space (bottom)}. Atomic Strategy (R1), and Transitive Rule (R2).}
    \label{fig:app_rules}
\end{figure}

We define two rules that transform the \ac{KG} into \KGFE: (i) Atomic tactics (Atomic Rule~\autoref{subsec:atomic}, addressing \kgap{1}), (ii) composite strategies (Transitive Rule~\autoref{subsec:transitive}, addressing \kgap{2}). Together, these rules form the Knowledge-Augmented Features Space in~\autoref{fig:app_rules}.

\subsubsection{Atomic Strategy Rule (R1)}\label{subsec:atomic}
The atomic rule enumerates a unique, representative strategy from each cluster of repetitive strategies. This rule ensures comprehensive coverage analysis of possible attack surface, including rare or legacy strategies and mechanisms that attackers may employ, which can potentially resurface in future attacks (as described in~\autoref{sec:our_approach}). Given clusters $\Pi = \{C_1, \dots, C_p\}$ from KG compression (\autoref{subsec:kg_compression}), we select one representative $S_j$ for each cluster that minimizes cumulative embedding of its name and description to all other strategies within the same cluster (\autoref{eq:cluster}):  
\begin{equation}
\mathcal{A} = \{ S_j \;:\; C_j \in \Pi \}.
\end{equation}

\subsubsection{Transitive Rule (R2)}\label{subsec:transitive}
The Transitive Rule captures composite strategies by identifying tactics that co-occur either within a single repository or across multiple repositories. This reflects how attackers may chain strategies, configuring multiple parameters jointly or combining tactics learned from different implementations. For instance, 
many TCP DoS tools allow enabling both \ttype{--ACK} and \ttype{--fragment}, yielding a \emph{Fragmented ACK} attack that merges flooding with fragmentation~\cite{vanhoef2021fragment}, or distributed DoS by combining 
\ttype{--IP-Spoofing} and \ttype{--Flood}.

We construct transitive edge  $E_{\mathrm{trans}} \subseteq \mathcal{A} \times \mathcal{A}$ as follows. First, \textit{Single-hop co-occurrence:} Two representative strategies $S_i, S_j \in \mathcal{A}$ are directly connected if $\exists r \in R, \exists s_1 \in C_i, \exists s_2 \in C_j$ such that $(s_1, r) \in E_{SR} \wedge (s_2, r) \in E_{SR}$. Second, \textit{Multi-hop chaining:} We compute the \textit{transitive closure} of these direct edges: if there exists a path $S_x \rightarrow S_i \rightarrow \cdots \rightarrow S_y$ through direct  edges, then $(S_x, S_y) \in E_{\mathrm{trans}}$.

For a representative strategy $S_x \in \mathcal{A}$, the set of transitively 
connected strategies is (illustrated in~\autoref{fig:app_rules}):
\begin{equation}
TC(S_x) = \{ S_y \in \mathcal{A} : (S_x, S_y) \in E_{\mathrm{trans}} \}.
\end{equation}

This construction exposes both direct (single-hop) and indirect (multi-hop) composites, expanding the feature space to cover variants that recombine known tactics in novel ways.

\subsection{\texorpdfstring{Knowledge-Augmented Feature Space \fullcirclednum{5}}{Knowledge-Augmented Features (5)}}\label{subsec:kg_augmented}
We translate the symbolic reasoning outputs (R1 and R2) into measurable network features that directly address the identified knowledge gaps. R1 produces atomic features addressing \kgap{1} (Out-of-Dimension 
Blindness), while R2 produces composite features with adaptive aggregation addressing \kgap{2} (Attack Strategy Aggregation Failure).

\paragraph{R1 $\rightarrow$ Atomic Strategy Features (Addressing \kgap{1})}  
The Atomic Rule enumerates representative strategies $\mathcal{A} = \{S_1, 
\dots, S_p\}$ from KG compression. Each atomic strategy $S_i \in \mathcal{A}$ maps to a measurable network feature $f_i$ via function $\phi_1$:
\begin{equation}
    \phi_1 : \mathcal{A} \rightarrow \mathcal{F}_{\phi_1}, \quad \phi_1(S_i) = f_i.
\end{equation}

For example, strategy \ttype{--window-size} maps to feature \emph{TCP window size}, while \ttype{--packet-size} maps to \emph{packet size}. These atomic features are extracted at the per-flow level (5-tuple: Source IP, Source Port, 
Destination IP, Destination Port, Protocol). By systematically enumerating dimensions from attack implementations, $\mathcal{F}_{\phi_1}$ incorporates behavioral dimensions absent from human-designed feature spaces.

\paragraph{R2 $\rightarrow$ Composite Features (Addressing \kgap{2})}  
The Transitive Rule identifies strategy pairs $(S_i, S_j)$ connected via $E_{\mathrm{trans}}$, indicating co-occurring tactics in composite attacks. For each such pair where $S_j \in TC(S_i)$, we construct a composite feature via function $\phi_2$:

\begin{equation}
\phi_2 : \{(S_i, S_j) : S_j \in TC(S_i)\} 
\;\rightarrow\; \mathcal{F}_{\phi_2}, 
\end{equation}
\begin{equation}
\phi_2(S_i, S_j) = f_i \otimes f_j
\end{equation}

where $\otimes$ denotes the aggregation operator (e.g., element-wise multiplication, joint counting). For instance, detecting \ttype{--ACK} and \ttype{--fragment} co-occurrence yields composite feature 
\ttype{tcp\_ack\_fragment\_count}.

Analysis of our KG reveals that 70\% of composite strategies involve IP spoofing (source IP manipulation), indicating distributed attack patterns where multiple sources target a single destination. This finding directly informs our aggregation design. To expose such M:1 attack patterns invisible in per-flow analysis (Observation 3), we compute composite features at \emph{destination-level aggregation}: statistics are aggregated across all flows sharing the same (Destination IP, Destination Port), regardless of source. Formally, for composite feature $f_i \otimes f_j$, we aggregate over all conversations $(H_{\text{src}}, H_{\text{dst}})$ where $H_{\text{dst}}$ is constant. 

\paragraph{Feature Value Extraction}
\label{subsec:feature_value_extraction}
For each feature $f_i \in \mathcal{F}_{\phi_1} \cup \mathcal{F}_{\phi_2}$, we compute four incremental statistics:  mean ($\mu$), standard deviation ($\sigma$), cumulative sum ($CS$), and sum of squared residuals ($SSR$). We employ Welford's online algorithm~\cite{welford1962note} and the damped windowing approach from~\cite{mirsky2018kitsune}, enabling constant-time updates with $O(1)$ memory per feature (details in~\autoref{app:fe_extract}). To summarize our two-level monitoring: 1) Channel-level:  Per-flow statistics identified by 5-tuple, capturing individual connections, and 2) Destination-level: Aggregated statistics for all traffic to a destination, identified by (Destination IP, Destination Port), capturing distributed attack patterns.

At time $t$, the complete \KGFE\ feature vector is:
\begin{equation}
    \bm{v}(t) = \mathcal{F}_{\phi_1} \cup \mathcal{F}_{\phi_2},
\end{equation}
where $\mathcal{F}_{\phi_1}$ contains atomic features addressing missing 
dimensions (\kgap{1}), and $\mathcal{F}_{\phi_2}$ contains destination-aggregated composite features addressing strategy combinations (\kgap{2}).

\section{Evaluation}\label{sec:eval}
%The design of \projectTitle's \KGFE\ aims to fulfill two objectives: 1) Detect attacks while maintaining low FPR, and 2) Improve detection effectiveness of current A-NIDS models compared to baseline feature spaces. 

We evaluate \projectTitle\ by addressing the following research questions:

\noindent\textbf{RQ1:} Does \projectTitle's \KGFE\ achieve effective attack detection while meeting the practical FPR requirements of realistic operational environments? (\autoref{subsec:low_for}) \newline
\noindent\textbf{RQ2:} Does \projectTitle's \KGFE\ improve generalization across different attack variants compared to other feature spaces? (\autoref{subsec:generalization})\newline
\textbf{RQ3:} Does \KGFE\ achieve computational efficiency suitable for online deployment rather than being limited to offline analysis? (\autoref{subsec:eval_eff}) \newline
\textbf{RQ4:} Can current machine learning techniques replace the need for \KGFE? (\autoref{subsec:eval_data_driven}) \newline
\textbf{RQ5:} How does each reasoning rule contribute to overall performance? (\autoref{subsec:ablation_study}) 

\paragraph{Experimental Methodology}
Our experimental evaluation follows best practices for A-NIDS assessment~\cite{apruzzese2023sokpragmaticassessmentmachine, arp2022and}.  We utilize 70\% of benign traffic for training, 10\% for validation and threshold tuning, and 20\% strictly held out for testing, using time-aware splitting. As established in~\autoref{sec:issues}, operationally practical FPR is $\leq$0.1\%. 

\paragraph{Prototype} We implemented a prototype of \projectTitle, covering three key components: \ac{KG} construction, Symbolic Reasoning, and the extraction of \KGFE\ from network traffic. Packet capture and preprocessing rely on \ttype{tshark}. The extraction module is designed as a plug-and-play component, supporting both offline analysis (\ttype{pcap} files) and online analysis, ensuring compatibility with a variety of \acp{A-NIDS}. For our prototype investigating TCP DoS, HTTP DoS, and Brute Force attacks, the KG is created from 7,853 repositories (details of the collection process are provided in~\autoref{app:source_acq}), and stored in \ttype{.graphml} format to support symbolic inference (for a summary of the extracted \projectTitle features, see~\autoref{app:feature_summary}). We extract 214 features in total, including flow-level metrics, temporal dynamics, and protocol-specific behaviors (comprehensive list in~\autoref{app:feature_summary}).

%bmk
\paragraph{Feature-Space Baselines} We use the three (CIC, SFS, and KIT) feature space baselines defined in~\autoref{sec:issues}.

\paragraph{Datasets} We evaluate \projectTitle on four datasets, each serving a distinct purpose in our evaluation methodology. We use two established benchmarks to assess performance across diverse deployment scenarios, as prior work shows that detection accuracy varies significantly by environment~\cite{nougnanke2025dataset}. The first benchmark is \acf{CIoT-23}\cite{neto2023ciciot2023}, which contains traffic from 105 IoT devices. The second is \acf{CIDS-17}\cite{sharafaldin2018toward}, which emulates an enterprise network with heterogeneous protocols. For \ac{CIDS-17}, we adopt the corrected labels from~\cite{9474286}. To evaluate generalization on attack variants absent from existing benchmarks, we collected CAP, which contains 9 variants targeting our identified knowledge gaps: out-of-dimension attacks exploiting  vulnerabilities disclosed  in 2025 (e.g. Nkiller2~\cite{CVE202432984}, HTTP Mal~\cite{NVD_CVE_2025_32472}), non-throughput mimicry attacks that overlap distributionally with benign traffic (e.g. Low-rate TCP~\cite{kuzmanovic2003low}), and parameter-variant attacks using identical tools with different configurations (e.g. SSH-Patator~\cite{lanjelot_patator} with P=0 vs P=1 from Concap~\cite{concap}). These parameter variants demonstrate that \projectTitle generalizes across configuration variations rather than overfitting to specific parameter settings. Complete details for dataset collection and processing are given in~\autoref{app:dataset details}.

\paragraph{A-NIDS Models} We evaluate each feature space, including that of \projectTitle, with three representative A-NIDS of varying complexity: (i) Gaussian Mixture Models (GMM)~\cite{an2022ensemble, zhang2020effective, amalapuram2022continual}\textemdash{}simple probabilistic baseline that models benign traffic as a mixture of Gaussians; (ii) Denoising Autoencoder (DA)~\cite{thakkar2021review}\textemdash{}a neural model that reconstructs benign traffic patterns, with anomalies identified via high reconstruction error; and (iii) Ensemble of Autoencoders (\KITML){}\footnote{%When reproducing Kitsune, 
Using the code from~\cite{mirsky2018kitsune}, we observed an exponential increase in feature-extraction runtime, as the number of packets grew. We identified the cause, implemented a patch, and verified that the original results were reproducible (see~\autoref{app:kitsune_patch}). All experiments are performed using the patched code for a fairer comparison.}~\cite{mirsky2018kitsune}\textemdash{}an ensemble trained on correlated feature subsets to exploit feature dependencies and learn richer representations from Kitsune~\cite{mirsky2018kitsune}.

\begin{table*}[t!]
\centering
\caption{
\textbf{Attack Detection at Low FPR.} 
F1-scores of \projectTitle and baselines. 
\textbf{Bold} marks the best feature space per detection model, 
\textcolor{Green}{\textuparrow} indicates \projectTitle improvements, 
and FPR values are presented in percentage (\%). 
Benign FPR$_{te}$ denotes the false positive rate on the hold-out benign-only test set. 
}

\label{tab:low_fpr_tuned}
\renewcommand{\arraystretch}{1.2}
\setlength{\tabcolsep}{3.0pt}
\scalebox{0.77}{
\begin{tabular}{l|r|r|r|r||r|r|r|r||r|r|r|r}
\hline
\multirow{2}{*}{\textbf{Attack}} 
& \multicolumn{4}{c||}{\textbf{GMM}} 
& \multicolumn{4}{c||}{\textbf{DA}}
& \multicolumn{4}{c}{\textbf{\KITML}} \\
\cline{2-13}
& SFS & CIC & KIT & \projectTitle 
& SFS & CIC & KIT & \projectTitle 
& SFS & CIC & KIT & \projectTitle \\
\hline
DoS-SYN (\ac{CIoT-23})
& 0.0439 & 0.0146 & 0.9173 & \textcolor{Green}{\textuparrow} \textbf{0.9834}
% DA
% 9939
& 0.0897 & 0.0593 & 0.9156 & \textcolor{Green}{\textuparrow} \textbf{0.9835}
% KIT
& 0.0930 & 0.0000 & \textbf{0.9990} & 0.9835 \\

DoS-TCP (\ac{CIoT-23})
& 0.1734 & 0.7509 & 0.8800 & \textcolor{Green}{\textuparrow} \textbf{1.0000}
&0.3273 & 0.3329 & 0.9786 & \textcolor{Green}{\textuparrow} \textbf{0.9999}

& 0.3357 & 0.0000 & 0.9988 & \textcolor{Green}{\textuparrow} \textbf{0.9999} \\
DoS-HTTP (\ac{CIoT-23})
% GMM
&0.6676 & 0.1453 & 0.0010 & \textcolor{Green}{\textuparrow} \textbf{0.9003} 
% DA
& 0.7277 & 0.0742 &\textbf{0.9991} & 0.9000
% KIT
& 0.6905 & 0.0256 & \textbf{0.9748} & 0.9001 \\
Brute Force (\ac{CIoT-23})
& 0.0000 & 0.0000 & 0.0000 & \textcolor{Green}{\textuparrow} \textbf{0.9727}
& 0.0000 & 0.0000 & 0.0000 & \textcolor{Green}{\textuparrow} \textbf{0.9750}
& 0.0000 & 0.0000 & 0.0000 & \textcolor{Green}{\textuparrow} \textbf{0.9750}  \\
DoS Hulk (\ac{CIDS-17})
% GMM
& 0.0228 & 0.1843 & 0.7637 & \textcolor{Green}{\textuparrow} \textbf{0.9178}
& 0.0406 & 0.2616 & 0.6317 & \textcolor{Green}{\textuparrow} \textbf{0.9677}
& 0.0270 & 0.5604 & 0.8641 & \textcolor{Green}{\textuparrow} \textbf{0.9679}
\\
DoS GoldenEye (\ac{CIDS-17})
% GMM
& 0.3609 & 0.2990 & 0.0150 & \textcolor{Green}{\textuparrow} \textbf{0.9934}
& 0.3122 & 0.4376 & 0.0005 & \textcolor{Green}{\textuparrow} \textbf{0.9981}
& 0.3052 & 0.7792 & 0.0000 & \textcolor{Green}{\textuparrow} \textbf{0.9953}
\\ 
Brute Force (\ac{CIDS-17})
& 0.0000 & 0.0000 & 0.0000 & \textcolor{Green}{\textuparrow} \textbf{0.9607}
& 0.0002 & 0.0000 & 0.0028 & \textcolor{Green}{\textuparrow} \textbf{0.4476}
& 0.0000 & 0.0000 & 0.0000 & \textcolor{Green}{\textuparrow} \textbf{0.2243} \\

\hline
\hline
\textbf{Benign $\text{FPR}_{te}$ (\ac{CIoT-23})}
& 0.0980\% & 0.0662\% & 0.0296\% & \textbf{0.0017}\%
& 0.0914\% & 0.1200\% & 0.0074\% & \textbf{0.0000}\%
& 0.0147\% & 0.1323\% & 0.0030\% & \textbf{0.0000}\%  \\

\textbf{Benign $\text{FPR}_{te}$ (\ac{CIDS-17})}
& 0.1203\% & 0.1021\% & 0.1186\% & \textbf{0.0137}\%
& 0.0971\% & 0.0994\% & 0.1465\% & \textbf{0.0000}\%
& 0.0957\% & 0.1047\% & 0.1967\% & \textbf{0.0076}\%  \\
\hline
\hline
\end{tabular}}
\end{table*}

\subsection{Attack Detection Performance at Low \ac{FPR}}\label{subsec:low_for}
The evaluation covers 3,384 scenarios per dataset (6,768 in total), combining 3 models, 94 hyperparameter configurations, 3 thresholds, and 4 feature spaces (i.e., $3\times94\times3\times4$, as detailed in ~\autoref{app:hyperparam}). For each model, we perform hyperparameter tuning to select the configuration that maximizes Recall while keeping the FPR at or below 0.1\%. Thus, every reported result corresponds to the best operating point of that model within this constraint, rather than to a single arbitrary threshold. This procedure reduces ``benchmark lottery'' effects~\cite{dehghani2021benchmark}, where apparent performance differences arise from accidental or suboptimal hyperparameter choices.

\autoref{tab:low_fpr_tuned} presents the results. Benign $\text{FPR}_{te}$ denotes the FPR on the holdout test set for the corresponding training environment. For instance, Benign $\text{FPR}_{te}$ (CIoT-23) refers to the value obtained when trained and tested on CIoT-23 benign data. Our features achieve higher F1-scores in most cases while consistently yielding the lowest false positive rate. KIT occasionally exceeds our F1-score but only at the cost of a higher FPR. The results also show the effect of background traffic complexity. KIT performs well in simpler environments such as CIoT-23, achieving an F1-score of 99.91\% for DA and 97.48\% for \KITML\ on HTTP DoS attacks. Its performance collapses in complex environments, for example, falling to nearly 0\% F1-score on the GoldenEye HTTP DoS variant in the enterprise-like CIDS-17 dataset. These findings support our claim in~\autoref{sec:issues}, confirming that Recall degradation under partial attack-benign overlap represents a systematic limitation across contemporary A-NIDS approaches. \KGFE's attack-relevant features enable discriminative separation of malicious traffic without false positive inflation, achieving 22.43-99.99\% F1-scores at lower FPR compared to baselines. This demonstrates that building feature space on attack knowledge resolves the fundamental FPR-Recall tradeoff under operational constraints.

\begin{center}
\fcolorbox{black}{gray!10}{%
  \parbox{0.9\linewidth}{%
    \textit{Takeaway 1:} \KGFE\ achieves attack detection F1-scores above 90\% in most cases while simultaneously operating at lower, operationally practical FPR levels than all baselines. 
  }
}
\end{center}

\subsection{Generalization on Attack Variants} \label{subsec:generalization}
To evaluate generalization of \projectTitle to variants and to highlight the limitations of existing approaches (\autoref{sec:issues}), we conduct experiments on diverse attack categories. The experiment is conducted using the same model parameters as in~\autoref{tab:low_fpr_tuned} (hence \ac{FPR} omitted as it remains unchanged). Results in~\autoref{tab:generalization} show that \projectTitle outperforms baselines on most variants, irrespective of the underlying anomaly model; in the few cases where \projectTitle does not outperform, the baselines result in a higher FPR (check with \autoref{tab:low_fpr_tuned}). Even a single two-layer DA trained with \KGFE\ achieves performance comparable to the ensemble of autoencoders. This suggests that well-chosen features can enable models to remain effective against new attack variants without increasing model complexity, an important property for deployment in resource-constrained environments.

\autoref{tab:generalization} also highlights how incorporating only narrow or incomplete attack knowledge into \ac{A-NIDS} feature design severely limits their capacity to detect unseen variants. First, \autoref{tab:generalization} shows that SFS and CIC consistently perform poorly on M-N endpoint aggregation. In the rare cases where their performance appears higher, we found that the attack traffic involved little variation in source IPs, reducing the scenario to what is effectively a 1-1 attack rather than a true M–N setting e.g., DDoS-ACK Fragmentation attack. Moreover, changing the model architecture does not remedy this limitation. In fact, SFS achieves higher scores on HTTP DoS across models and SYN DoS performance remains fixed at 0\% F1-score, indicating that feature space, not model type, is the primary determinant of detection capability. In contrast, KIT aggregates features at the source level, giving the model a global view that yields stronger results on M–N attacks.

% Second, \autoref{tab:generalization} confirms that existing feature spaces are highly susceptible to Out-of-Dimension blindness, with contemporary approaches often collapsing to 0\% F1-score across all models. This further underscores that increasing model type alone does not improve detection when the feature space fails to capture attack-relevant dimensions. An exception arises with CIC on the Nkiller2 variant. CIC's partial detection of Nkiller2 (77.92-86.96\%) does not contradict Observation 2 but rather validates our thesis. CIC succeeds only because its feature space coincidentally includes TCP window statistics, the exact dimension Nkiller2 exploits. This represents fragile, attack-specific success dependent on whether feature designers anticipated this particular strategy. In contrast, SFS and KIT completely fail (0\% F1-score) despite identical model architectures, confirming that out-of-dimension blindness is a feature space property, not a model limitation.

Second, \autoref{tab:generalization} confirms that existing feature spaces are highly susceptible to Out-of-Dimension blindness, with contemporary approaches often collapsing to 0\% F1-score across all models. This further underscores that increasing model type alone does not improve detection when the feature space fails to capture attack-relevant dimensions. An exception arises with CIC on the Nkiller2 variant. CIC's partial detection of Nkiller2 (77.92-86.96\%) does not contradict Observation 2 but rather validates our thesis. CIC succeeds only because its feature space includes TCP window statistics, the exact dimension Nkiller2 exploits. This finding further confirms that encoding attack dynamics into the feature space is critical for effective detection. In contrast, SFS and KIT completely fail (0\% F1-score) despite identical model architectures, confirming that out-of-dimension blindness is a feature space property, not a model limitation.

Similarly, models lacking attack-relevant semantics fail to generalize to non-throughput mimicry attacks. CIC attains an F1-score of 86.96\% on Low-rate TCP only because its features include several TCP-level indicators (e.g., SYN flag counts) directly tied to the attack’s mechanics. This attack exploits carefully timed bursts of SYN packets synchronized with the TCP retransmission-timeout mechanism~\cite{kuzmanovic2003low}, and CIC’s feature space captures these semantics. These findings emphasize that generalization hinges less on architecture than on whether the feature space encodes attack-relevant semantics. By embedding this knowledge, \projectTitle overcomes these limitations and achieves strong results at low \ac{FPR}.

\begin{center}
\fcolorbox{black}{gray!10}{%
  \parbox{0.9\linewidth}{%
    \textit{Takeaway 2:} Baselines achieved 0\% on out-of-dimension attacks and M-N distributed attacks, 
    confirming Observations 2-3 and both knowledge gap categories. \KGFE\ achieved 76.99-100\% on these attacks, validating that mining attack implementations addresses out-of-dimension blindness (\kgap{1}) and aggregation failure (\kgap{2}).
  }
}
\end{center}

\begin{table*}\small
\centering
\caption{
\textbf{Generalization on Attack Variants.} 
F1-scores of \projectTitle and baselines. 
\textbf{Bold} marks the best feature space per detection model, 
\textcolor{Green}{\textuparrow} indicates \projectTitle improvements, 
and FPR values are presented in percentage (\%).  
}
\label{tab:generalization}
\renewcommand{\arraystretch}{1.2}
\setlength{\tabcolsep}{4pt}
\scalebox{0.75}{
\begin{tabular}{c|l|r|r|r|r|r|r|r|r|r|r|r|r} 
\hline
\multicolumn{1}{c|}{\multirow{2}{*}{\textbf{Category}}} & 
\multicolumn{1}{c|}{\multirow{2}{*}{\textbf{Attack}}} & 

% --- GMM ---
\multicolumn{4}{c|}{\textbf{GMM}} & 

% --- DA ---
\multicolumn{4}{c|}{\textbf{DA}} & 

% --- KIT-ML ---
\multicolumn{4}{c}{\textbf{\KITML}} \\ 
\cline{3-14}

& \multicolumn{1}{l|}{}  
& SFS  & CIC    & KIT    &   \projectTitle   % <-- GMM columns
& SFS  & CIC    & KIT    &   \projectTitle   % <-- DA columns
& SFS  & CIC    & KIT    &   \projectTitle   % <-- KIT-ML columns
\\ 
\hline

% =========================================================
% OUT-OF-DIMENSION
% =========================================================
\multirow{4}{*}{Out-of-Dimension} 
& HTTP Mal (CAP)        & 0.0000 & 0.0000 & 0.0000 & \textcolor{Green}{\textbf{\textuparrow}} \textbf{0.9844} 
                        & 0.0000 & 0.0000 & 0.0000 & \textcolor{Green}{\textbf{\textuparrow}} \textbf{0.9985} 
                        & 0.0000 & 0.0000 & 0.0000 & \textcolor{Green}{\textbf{\textuparrow}} \textbf{0.9985} \\
& HTTP Overflow (CAP)   & 0.0000 & 0.0000 & 0.0000 & \textcolor{Green}{\textbf{\textuparrow}} \textbf{0.9817} 
                        & 0.0000 & 0.0000 & 0.0000 & \textcolor{Green}{\textbf{\textuparrow}} \textbf{0.9770} 
                        & 0.0000 & 0.0000 & 0.0000 & \textcolor{Green}{\textbf{\textuparrow}} \textbf{0.9800} \\
& Nkiller2 (CAP)        & 0.0000 & 0.8696 & 0.0000 & \textcolor{Green}{\textbf{\textuparrow}} \textbf{0.8841} 
                        & 0.0000 & 0.7792 & 0.0000 & \textcolor{Green}{\textbf{\textuparrow}} \textbf{0.9190} 
                        & 0.0000 & 0.8451 & 0.0000 & \textcolor{Green}{\textbf{\textuparrow}} \textbf{0.9276} \\
& Brute Force P=0 (Con) 
                        & 0.0000 & 0.0000 & 0.0002 & \textcolor{Green}{\textbf{\textuparrow}} \textbf{0.9998} 
                        & 0.0000 & 0.0000 & 0.0000 & \textcolor{Green}{\textbf{\textuparrow}} \textbf{0.9998} 
                        & 0.0000 & 0.0000 & 0.0000 & \textcolor{Green}{\textbf{\textuparrow}} \textbf{0.9998} \\
\hline

% =========================================================
% NON-THROUGHPUT
% =========================================================
\multirow{8}{*}{Non-Throughput} 

& Fiberfox AVB (CAP)    & \textbf{0.9622} & 0.0000 & 0.0000 & 0.8523
                        & 0.0042 & 0.0000 & 0.0000 & \textcolor{Green}{\textbf{\textuparrow}} \textbf{0.7649}
                        & 0.0042 & 0.0000 & 0.0000 & \textcolor{Green}{\textbf{\textuparrow}} \textbf{0.8638} \\
& Fiberfox BYPASS (CAP) & 0.3205 & 0.0000 & 0.2786 & \textcolor{Green}{\textbf{\textuparrow}} \textbf{0.9185}
                        & 0.0000 & 0.5509 & 0.0045 & \textcolor{Green}{\textbf{\textuparrow}} \textbf{0.8576}
                        & 0.0000 & 0.0000 & 0.0030 & \textcolor{Green}{\textbf{\textuparrow}} \textbf{0.8431} \\
& Fiberfox GET (CAP)    & 0.0321 & 0.1623 & 0.0064 & \textcolor{Green}{\textbf{\textuparrow}} \textbf{0.9766}
                        & 0.2770 & 0.3205 & \textbf{0.9957} & 0.9742
                        & 0.0298 & 0.1619 & \textbf{0.9997} & 0.9742 \\
& Brute Force P=1 (Con) 
                        & 0.0000 & 0.0000 & 0.0000 & \textcolor{Green}{\textbf{\textuparrow}} \textbf{0.9994}
                        & 0.0000 & 0.0000 & 0.0000 & \textcolor{Green}{\textbf{\textuparrow}} \textbf{0.9994}
                        & 0.0000 & 0.0000 & 0.0000 & \textcolor{Green}{\textbf{\textuparrow}} \textbf{0.9994} \\
& DDoS-Slowloris (CIoT-23) 
                        & 0.0287 & 0.0027 & 0.0001 & \textcolor{Green}{\textbf{\textuparrow}} \textbf{0.8114}
                        & 0.0000 & 0.0000 & 0.0015 & \textcolor{Green}{\textbf{\textuparrow}} \textbf{0.7548}
                        & 0.0302 & 0.0001 & 0.5198 & \textcolor{Green}{\textbf{\textuparrow}} \textbf{0.7560} \\
& DoS-Slowhttptest (CIDS-17) 
                        & 0.0031 & 0.0000 & 0.0000 & \textcolor{Green}{\textbf{\textuparrow}} \textbf{0.9214}
                        & 0.0002 & 0.3117 & 0.0018 & \textcolor{Green}{\textbf{\textuparrow}} \textbf{0.9652}
                        & 0.0000 & 0.0281 & 0.0000 & \textcolor{Green}{\textbf{\textuparrow}} \textbf{0.9647} \\
& DoS-SlowLoris (CIDS-17) 
                        & 0.0568 & 0.0000 & 0.0031 & \textcolor{Green}{\textbf{\textuparrow}} \textbf{0.9803}
                        & 0.0000 & 0.0131 & 0.0028 & \textcolor{Green}{\textbf{\textuparrow}} \textbf{0.9714}
                        & 0.0000 & 0.0000 & 0.0000 & \textcolor{Green}{\textbf{\textuparrow}} \textbf{0.9818} \\
& Low rate TCP (CAP)    & 0.0000 & \textbf{0.8696} & 0.0000 & 0.7699
                        & 0.0000 & 0.7792 & 0.0000 & \textcolor{Green}{\textbf{\textuparrow}} \textbf{0.9194}
                        & 0.0000 & 0.8451 & 0.0000 & \textcolor{Green}{\textbf{\textuparrow}} \textbf{0.9277} \\  
\hline

% =========================================================
% M-N END-POINT
% =========================================================
\multirow{9}{*}{Composite}  
& DDoS-HTTP (CIoT-23)    & 0.8360 & 0.0022 & 0.0002 & \textcolor{Green}{\textbf{\textuparrow}} \textbf{0.9696}
                        & 0.8502 & 0.0101 & 0.2909 & \textcolor{Green}{\textbf{\textuparrow}} \textbf{0.9476}
                        & 0.8418 & 0.0033 & 0.8506 & \textcolor{Green}{\textbf{\textuparrow}} \textbf{0.9476} \\
& DDoS-SYN (CIoT-23)     & 0.0000 & 0.9099 & 0.1292 & \textcolor{Green}{\textbf{\textuparrow}} \textbf{0.9979}
                        & 0.0000 & 0.0004 & 0.9620 & \textcolor{Green}{\textbf{\textuparrow}} \textbf{0.9979}
                        & 0.0000 & 0.0000 & 0.9873 & \textcolor{Green}{\textbf{\textuparrow}} \textbf{0.9979} \\
& DDoS-TCP (CIoT-23)     & 0.0000 & 0.0000 & 0.4294 & \textcolor{Green}{\textbf{\textuparrow}} \textbf{0.9997}
                        & 0.0000 & 0.0000 & 0.9992 & \textcolor{Green}{\textbf{\textuparrow}} \textbf{0.9997}
                        & 0.0000 & 0.0000 & 0.9986 & \textcolor{Green}{\textbf{\textuparrow}} \textbf{0.9999} \\
& DDoS-PSHACK (CIoT-23)  & 0.0000 & 0.2405 & 0.0728 & \textcolor{Green}{\textbf{\textuparrow}} \textbf{0.9778}
                        & 0.0000 & 0.0008 & 0.9656 & \textcolor{Green}{\textbf{\textuparrow}} \textbf{0.9753}
                        & 0.0000 & 0.0078 & 0.9831 & \textcolor{Green}{\textbf{\textuparrow}} \textbf{0.9772} \\
& DDoS-RSTFIN (CIoT-23)  & 0.0000 & 0.7057 & 0.5166 & \textcolor{Green}{\textbf{\textuparrow}} \textbf{1.0000}
                        & 0.0000 & 0.0000 & 0.9983 & \textcolor{Green}{\textbf{\textuparrow}} \textbf{1.0000}
                        & 0.0000 & 0.0000 & 0.9950 & \textcolor{Green}{\textbf{\textuparrow}} \textbf{1.0000} \\
& DDoS-SynIP (CIoT-23)   & 0.0000 & 0.0000 & 0.9992 & \textcolor{Green}{\textbf{\textuparrow}} \textbf{1.0000}
                        & 0.0000 & 0.8000 & 0.9999 & \textcolor{Green}{\textbf{\textuparrow}} \textbf{1.0000}
                        & 0.0000 & 0.8001 & 0.9999 & \textbf{1.0000} \\
& DDoS-ACK Frag (CIoT-23) 
                        & \textbf{0.9768} & 0.8534 & 0.0006 & 0.9438
                        & \textbf{0.9803} & 0.8442 & 0.5375 & 0.9438
                        & \textbf{0.9802} & 0.8237 & 0.9476 & 0.9438 \\
& Fiberfox SLOW (CAP)    & 0.0000 & 0.1828 & 0.2637 & \textcolor{Green}{\textbf{\textuparrow}} \textbf{0.9825}
                        & 0.0000 & 0.2514 & 0.9959 & \textcolor{Green}{\textbf{\textuparrow}} \textbf{0.9796}
                        & 0.0000 & 0.1610 & 0.9997 & \textcolor{Green}{\textbf{\textuparrow}} \textbf{0.9796} \\
& Fiberfox STRESS (CAP)  & 0.9312 & 0.2492 & 0.0000 & \textcolor{Green}{\textbf{\textuparrow}} \textbf{0.9605}
                        & 0.9157 & 0.9914 & 0.9869 & \textcolor{Green}{\textbf{\textuparrow}} \textbf{0.9291}
                        & 0.8814 & 0.9930 & \textbf{0.9992} & 0.9295 \\
\hline

\end{tabular}}
\end{table*}

\subsection{Computational Efficiency}\label{subsec:eval_eff}
High detection performance is insufficient if feature space extraction creates computational bottlenecks in operational settings. We evaluate \KGFE's computational efficiency to verify it provides a practical solution rather than a theoretically effective but operationally infeasible approach. We evaluated computational efficiency along two dimensions: 1) feature extraction runtime, and 2) model inference time. We compare against Kitsune~\cite{mirsky2018kitsune}, which explicitly claims deployability as an online NIDS. Since the \ttype{C++} implementation of Kitsune is not publicly available, we use the released Python version~\cite{KitsunePy}. In addition, our evaluation is constrained to a single model rather than the full ensemble. In Kitsune, the Feature Mapper (FM) allocates the number of denoising autoencoders dynamically based on feature correlations (see~\cite{mirsky2018kitsune} for details). To ensure a fair and unified comparison, we evaluate only the single underlying model for both evaluated feature spaces. \projectTitle is also implemented in Python, which avoids bias from implementations in different programming language. All experiments were conducted on an Ubuntu 20.04 VM (2.6\,GHz CPU, single thread, 4\,GB RAM).

\begin{figure}[t]
    \centering
    \includegraphics[width=0.9\linewidth]{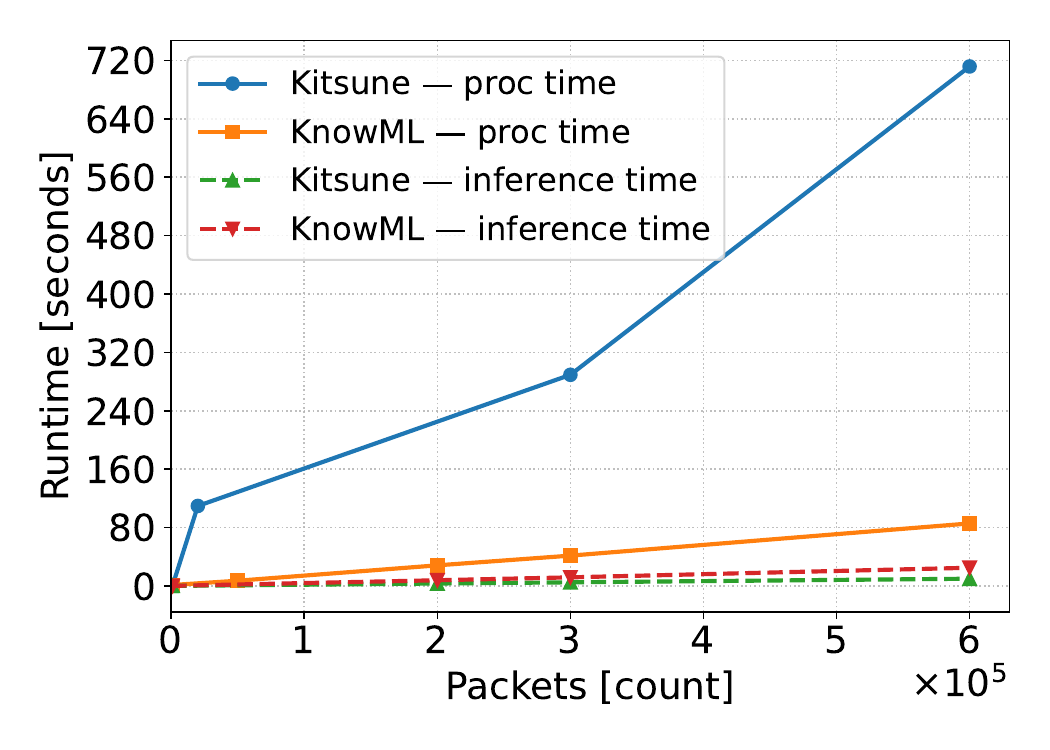}
    \caption{\textbf{Computational Efficiency.} \projectTitle vs. Kitsune~\cite{mirsky2018kitsune}: (a) feature extraction and (b) inference time. X-axis: input packets; Y-axis: runtime (s) with DA model.
    }
    \label{fig:efficiency}
\end{figure}

\autoref{fig:efficiency} presents the results. Kitsune scales poorly with traffic volume, processing 1k packets required 0.42s, while 600k packets required 711.85s, showing a non-linear runtime increase. By contrast, \projectTitle processed 600k packets in 85.59s ($\approx$7,010 pps), despite having more than twice the feature dimensionality (214 vs. 100\footnote{The  original paper~\cite{mirsky2018kitsune} claims 115 features, whereas its  artifact 100~\cite{KitsunePyIssue12}.}). This throughput is comparable to the 7,500 pps reported for Kitsune’s \ttype{C++} implementation on a faster 3.6\,GHz CPU~\cite{mirsky2018kitsune}. Given that \ttype{C++} implementations are typically 8–29× faster than Python under identical conditions~\cite{lion2022investigating}, a \ttype{C++} version of \projectTitle\ would likely surpass Kitsune in feature extraction speed.  In terms of inference, \projectTitle\ is slower due to the higher feature dimensionality, achieving $\approx$24,836 pps compared to Kitsune’s $\approx$61,166 pps. However, this overhead is offset by its more efficient extraction pipeline, supporting the suitability of \projectTitle's feature space for practical deployment.

\begin{center}
\fcolorbox{black}{gray!10}{%
  \parbox{0.9\linewidth}{%
    \textit{Takeaway 3:} The \KGFE\ feature space is practically obtainable in operational settings and does not introduce prohibitive computational overhead, achieving feature extraction efficiency comparable to SotA A-NIDS online systems. 
  }
}
\end{center}

\subsection{Comparing with Data-Driven Approaches}\label{subsec:eval_data_driven}
We evaluate whether purely data-driven frameworks for generalization can substitute for the \KGFE. This experiment tests if advances in representation learning alone are sufficient, or whether semantic features remain necessary. We focus on two recent methods that represent strong SotA approaches. AOC-IDS\cite{zhang2024aoc} employs a Cluster Repelling Contrastive loss within an autoencoder and reports over 91\% zero-day detection. Trident\cite{zhao2024trident} is an incremental learning framework that operates on existing feature spaces to improve generalization, using tScissors for outlier thresholding and tMagnifier for clustering-based class discovery. Trident and AOC-IDS have demonstrated improved generalization on existing models~\cite{whisper,hypervision,barradas2021flowlens}, making them suitable as competitive baselines for evaluating whether data-driven generalization alone can replace semantic knowledge.

We applied both methods to the KIT, chosen because it performs well on M–N endpoint attacks but shows poor results on Out-of-Dimension and Non-Throughput variants; specifically, we focus on scenarios in which KIT performed 0\% across all three detection models,  to test whether advanced data-driven learning can compensate for KIT’s limitations. Due to computational costs (even A100 GPUs with 80GB memory were insufficient), we trained on a sub-sampled dataset that remained 125 times larger than the original AOC-IDS dataset and comparable in size to Trident, with hold-out test data to avoid data snooping. Results in ~\autoref{tab:data_driven_study} show that both methods fail to generalize to attack variants. In the few cases where performance improves, it comes at the cost of exceptionally high false positive rates, for example, 20.67\% with Trident. These findings reveal that machine learning techniques such as contrastive and incremental learning cannot substitute for knowledge-based feature spaces, failing to generalize under operationally required FPR constraints.

\begin{center}
\fcolorbox{black}{gray!10}{%
  \parbox{1\linewidth}{%
    \textit{Takeaway 4:} Purely data-driven generalization frameworks, including contrastive and incremental learning, provide limited improvements under operational FPR constraints and do not replace the knowledge-augmented feature space.
  }
}
\end{center}

\begin{table}[t!]
\centering
\caption{\textbf{Comparing with Data-Driven Approaches.} Per-attack F1-scores of AOC-IDS, Trident, Kitsune, and \projectTitle, along with each approach’s FPR on benign traffic on the same sub-sampled dataset.}
\label{tab:data_driven_study}
\setlength{\tabcolsep}{8pt}
\renewcommand{\arraystretch}{1.25}
\scalebox{0.85}{
\setlength{\tabcolsep}{3pt}
\begin{tabular}{@{} l cccc @{}}
\toprule
\textbf{Attack} & \multicolumn{4}{c}{\textbf{F1-score}} \\
\cmidrule(lr){2-5}
 & \makecell{\textbf{AOC-IDS~\cite{zhang2024aoc}}} 
 & \makecell{\textbf{Trident~\cite{zhao2024trident}}}
 & \makecell{\textbf{Kitsune}}
 & \makecell{\textbf{\projectTitle}} \\
\midrule
Low rate TCP (SIM)        & 0.0000 & 0.0000 & 0.0000 & \textbf{0.9247} \\
Fiberfox AVB (CAP)        & 0.1659 & 0.7952 & 0.0000 & \textbf{0.8702} \\
Fiberfox BYPASS (CAP)     & 0.0536 & 0.9302 & 0.0000 & \textbf{0.9565} \\
HTTP Overflow (SIM)       & 0.0000 & 0.0000 & 0.0000 & \textbf{0.9703} \\
Nkiller2 (SIM)            & 0.0000 & 0.0000 & 0.0000 & \textbf{0.9348} \\
Brute Force P=1 (CAP) & 0.0180 & 0.9744 & 0.0000 & \textbf{1.0000} \\
\midrule
\multicolumn{1}{l}{\textit{Benign FPR}$_{te}$} 
 & \textbf{0.0043} & \cellcolor{red!30}\textbf{0.2067} & \textbf{0.0011} & \textbf{0.0000} \\
\bottomrule
\end{tabular}
}
\end{table}

\subsection{Ablation Study}\label{subsec:ablation_study}
We perform an ablation study on rule-derived features from~\autoref{subsec:symbolic_reasoning}. Using the DA model, we evaluate each rule set independently and in combination (\autoref{tab:ablation_study}). Atomic Features (R1) achieve strong performance across categories (average 79.19\% F1), excelling on out-of-dimension attacks (93.73\%) by directly encoding strategy-specific dimensions extracted from attack parameters. Composite Features (R2) perform best on M-N endpoint attacks (96.43\%) where multi-source behavior requires aggregation across temporal and spatial dimensions, and attacks that employ composite strategies e.g., R2 features achieve $94.38\%$ on DDoS ACK-Fragmentation attack compared to 31.36\% F1-score for R1.  Overall, each rule contributes complementary strengths. R1 ensures broad strategy coverage, and R2 captures multi-step combinations. Both rule sets are essential for generalization, as their integration outperforms either rule alone, demonstrating that effective threat knowledge embedding requires both atomic primitives and their compositions.

\begin{center}
\fcolorbox{black}{gray!10}{%
  \parbox{\linewidth}{%
    \textit{Takeaway 5:} Features derived from Atomic Rules (R1) and Transitive Rules (R2) provide complementary benefits, with each covering different attack variants. Their combination yields the best overall performance, indicating that effective threat knowledge embedding requires both fine grained strategy primitives and their aggregated forms.
  }
}
\end{center}

\section{Discussion}\label{sec:discussion}

\paragraph{Manual Validation for Feature Mapping}
\projectTitle provides a general framework for structural strategy analysis; in this work, we instantiate it with a concrete feature space designed to capture the identified strategies and their corresponding attack behaviors. Our experimental results indicate that this instantiation is effective, yet the framework is not tied to a single mapping, and we encourage future work to explore alternative feature designs within \projectTitle. Importantly, the aim of \projectTitle is not to replace human experts. An experienced analyst may be equally capable of identifying relevant attack strategies. Rather, \projectTitle automates the labor-intensive and fatigue-prone tasks (collecting, annotating, organizing, and analyzing attack implementations) that are difficult to perform manually at scale (7,853 repositories in our evaluation).

We intentionally retain a human in the loop to incorporate domain-specific expertise in feature mapping, particularly when translating abstract strategies into concrete network-level measurements and aggregation schemes as described in \autoref{subsec:kg_augmented}.  
This is a one-time effort per attack family that subsequently enables detection of numerous variants. In contrast, traditional approaches require expert analysis for each individual variant. Our evaluation demonstrates that a single knowledge-guided feature space, once validated, generalizes across multiple variants (\autoref{tab:generalization}), amortizing the upfront expert effort. 
As LLMs continue to advance, they may assist in automating this step. We leave a systematic exploration of this direction to future work. 

% \projectTitle provides a general framework for structural strategy analysis; in this work, we instantiate it with a concrete feature space designed to capture the identified strategies and their corresponding attack behaviors. Our experimental results indicate that this instantiation is effective, yet the framework is not tied to a single mapping, and we encourage future work to explore alternative feature designs within \projectTitle. 

% We intentionally retain a human in the loop to incorporate domain-specific expertise in feature mapping, particularly when translating abstract strategies into concrete network-level measurements and aggregation schemes as described in \autoref{subsec:kg_augmented}.  
% %
% This is a one-time effort per attack family that subsequently enables detection of numerous variants. In contrast, traditional approaches require expert analysis for each individual variant. Our evaluation demonstrates that a single knowledge-guided feature space, once validated, generalizes across multiple variants (\autoref{tab:generalization}), amortizing the upfront expert effort. 
% %
% As LLMs continue to advance, they may assist in automating this step. We leave a systematic exploration of this direction to future work. 

\begin{table}[t!]\small
\centering
\caption{\textbf{Ablation Study}. Average F1-scores on the DA model when using features derived from individual rules.
A tick (\cmark) denotes inclusion of features from the corresponding rule, while a cross (\xmark) denotes their removal.}
\label{tab:ablation_study}

\scalebox{0.85}{%
\begin{tabular}{|c|c|c|c|}
\hline
\textbf{Category}
  & \makecell{\textbf{R1} \\ (\cmark, \xmark)}
  & \makecell{\textbf{R2} \\ (\xmark, \cmark)}
  & \makecell{\textbf{R1,R2} \\ (\cmark, \cmark)} \\
\hline
Out-of-Dimension     & 0.9373 & 0.7834 & 0.9761 \\
\hline
Non-Throughput-based & 0.7706 &  0.4542   & 0.9219 \\
\hline
Composite         & 0.6678 & 0.9643 & 0.9815 \\
\hline\hline
Benign $\text{FPR}_{te}$ & 0.0006 & 0.0000 & 0.0000 \\
\hline
\end{tabular}%
}
\end{table}

\paragraph{Choice of Attack Families} In this paper, we focus on TCP DoS, HTTP DoS, and SSH Brute Force. These families are common in real-world deployments~\cite{crowdstrike_common_attacks} and widely studied~\cite{staicu2018freezing, harris1999tcp, javed2013detecting}, allowing us to demonstrate that existing solutions fail to generalize even within well-studied families. While it is easier to show gaps in \ac{A-NIDS} using obscure or uncommon attacks, doing so with well-known and heavily researched variants demonstrates the practical utility of our approach. 
%
%Our objective is not to limit the framework to these families, but to establish a general methodology., 
\projectTitle enables the systematic enumeration of attack strategies and the examination of relationships between them. This design makes it extensible to the broader threat landscape. While our evaluation uses GitHub-sourced strategies, the KG can be enriched with proprietary threat intelligence, incident response data, or frameworks such as MITRE ATT\&CK. Organizations can augment \projectTitle with internal knowledge sources to address environment-specific threats not present in public repositories.

\paragraph{Adversarial ML Attacks}
%Considerations on Sophisticated Adversaries}
We evaluate \projectTitle in the context of generalization of detection tasks with attack variants, and do not directly assess its resilience to adversarial ML.
Since \projectTitle{} distills behavioral features, we expect knowledge-augmented features to be more robust to adversarial manipulation, as attackers must satisfy problem-space constraints~\cite{pierazzi2020intriguing}.
We plan to explore how knowledge-augmented systems may mitigate adversarial threats in future work. 

 \paragraph{Other \ac{A-NIDS} Feature-Space Baselines} 
Other feature-space baselines include FlowLens~\cite{barradas2021flowlens}, Whisper~\cite{whisper}, and HyperVision~\cite{hypervision}. These methods extract few packet-level features such as size and inter-packet timing (throughput-based), then transform them through quantization, Fourier analysis, or flow-level statistics. They function as data-driven feature enhancement techniques. Our results show that Kitsune, which monitors 100 throughput-based statistics, cannot detect Out-of-Dimension or Non-Throughput attacks even when combined with SotA enhancement or incremental learning frameworks (\autoref{subsec:eval_data_driven}). This suggests other approaches face the same limits. That being said, HyperVision, which uses connection degree for detection, might detect M-N endpoint attacks but would suffer from the same constraint as Kitsune in detecting the other two categories of attacks. It is important to note that these systems were designed for efficiency and Tbps-scale deployment, sometimes on programmable switches, and intentionally use very few features, whereas \projectTitle focuses on semantic generalization across attack variants. In future work, we plan to systematically study feature spaces from different paradigms, particularly with focus on high-speed networks. 

% We plan to explore approaches to reduce feature dimensionality, further abstract behaviors, embed attack knowledge directly into models, and neuro-symbolic approaches to balance generalization with efficiency as future work.

\paragraph{Detecting Zero-Days}
\projectTitle's \ac{KG} models diverse attack instances in order to distill generalizable methodology patterns rather than overfitting to specific cases. As a result, a model using the resulting \KGFE\ detects variants that extend or combine known strategies but cannot anticipate entirely new vulnerabilities outside this space. When such a vulnerability emerges, however, integrating it requires only a constant-time update to the \ac{KG} (e.g., adding a new edge), enabling the framework to evolve quickly as new threat intelligence becomes available.

\paragraph{Cost of Knowledge Graph Construction}
Constructing the KG remains scalable and cost-efficient even for large codebases. For instance, extracting data and text embeddings for TCP DoS across 2,333 repositories using GPT-4o-mini cost approximately US\$13. Beyond monetary cost, \projectTitle substantially reduces human effort. We estimate that manual expert analysis of TCP DoS, HTTP DoS, and Brute-Force families would require approximately 164 workdays (7,853 repositories at 10 minutes each over 8-hour workdays). \projectTitle automates the bulk of this effort through large-scale analysis of attack implementations.

\section{Related Work}\label{sec:related_work}

\paragraph{KG Applications in Security}
\acp{KG} have gained traction in cybersecurity, with a range of applications from detection to knowledge management~\cite{zou2020survey}. The most relevant to our work are those that directly integrate KGs into detection pipelines. For example, Yang et al.~\cite{yang2022enhanced} construct a KG from NSL-KDD feature co-occurrence and embed it into a CNN-BiLSTM. Their approach enhances raw features through data-driven correlations, whereas \projectTitle introduces an attack-implementation-based feature space grounded in domain knowledge, rather than manipulations of existing features. Another approach is  KnowGraph~\cite{zhou2024knowgraph}, which models buyer–seller interactions as a KG for anomaly detection, learning embeddings via GNNs and refining predictions with expert-provided probabilistic rules. However, KnowGraph depends on external corrections (expert-defined rules) and does not expand the knowledge base. By contrast, \projectTitle embeds foundational attack knowledge directly into the feature space, allowing models to capture the broader attack landscape to improve generalization.

\paragraph{Automated Feature Engineering in Security} 
Automated feature engineering in security follows two directions. Representation learning derives latent features from data to improve detection~\cite{wang2022representation, 10646725, AOC-IDS}. Feature mining uses external knowledge to refine an existing feature space, as in FeatureSmith~\cite{zhu2016featuresmith}, which enhances Drebin~\cite{arp2014drebin} by selecting useful features. \projectTitle instead expands the feature space with attack-relevant features grounded in implementation details, creating new semantics rather than filtering existing ones. This approach outperforms prior methods focused only on feature enhancement (\autoref{subsec:eval_data_driven}).

\paragraph{Exploring Limitations of ML-NIDS} 
Recent critiques emphasize two issues: 1) dataset inadequacies~\cite{apruzzese2023sokpragmaticassessmentmachine, flood2024bad, 9474286, nougnanke2025dataset} and 2) flawed ML design practices~\cite{arp2022and, d2022establishing}. Flood et al.~\cite{flood2024bad} describe ``bad smells'' in benchmark datasets, such as \textit{poor diversity}, which produces highly dependent features. Such flaws prevent models from learning meaningful representations, limit generalization, and inflate reported performance. In contrast, we analyze limitations from a semantic perspective. We reason about how well features capture attack semantics, i.e., their ability to generalize, how this affects their ability to separate benign from malicious traffic, and how it influences the number of false positive alarms.

\section{Conclusion}
This paper identifies fundamental gaps in SotA feature spaces used by \acf{A-NIDS}, showing that they are insufficient for capturing attack variants.
To address this gap, we proposed \projectTitle, a framework that, given attack families of interest, builds  a KG of attack strategies and applies symbolic reasoning to derive a corresponding Knowledge-Augmented Feature Space (\KGFE) grounded in attack semantics. 

Our evaluation on established benchmarks and new attack variants revealed two categories of knowledge gaps in existing approaches. We demonstrated that \projectTitle closes these gaps, achieving effective generalization while maintaining low false positive rates. These results highlight the importance of incorporating attack-relevant semantics into the feature space of A-NIDS. The findings also have broader implications. They encourage the exploration of semantic reasoning in other security applications and motivate the development of adaptive defenses that evolve with attacker strategies.

% \section*{Ethical Considerations}
% We conducted our research ethically, with the principle of beneficiance in mind. To the best of our knowledge, this work and its experiments raise no ethical concerns. We did not collect any traffic without user consent. We primarily used two public datasets (\ac{CIDS-17} and \ac{CIoT-23}), and then generated attack variants dataset, CAP via traffic generation on controlled lab machines, on which launching the attacks caused no harm. For attack implementations, we crawled publicly available repositories. Our approach and methodology follows established best practices in network security research, consistent with prior work~\cite{sommer2010outside,flood2024bad,apruzzese2023sokpragmaticassessmentmachine, arp2022and}.

\section*{Ethical Considerations}
We conducted our research ethically, with the principle of beneficence in mind. To the best of our knowledge, neither the work nor its experiments raise ethical concerns, no traffic was collected without user consent, and the CAP dataset was generated on controlled lab machines where the attacks caused no harm.

\section*{Acknowledgments}
This work was supported by: the UKRI Centre for Doctoral Training in Safe and Trusted Artificial Intelligence (EP/S023356/1), the UK EPSRC Grant no. EP/X015971/2, a  Google Academic Research Award (GARA), and Google ASPIRE awards. We thank the anonymous reviewers for their valuable feedback, and Giovanni Apruzzese for providing several attack variants used in this study.

\bibliographystyle{plain}
\bibliography{references}

\section{Data Availability}
Source code is available at: \url{https://github.com/s2labres/KnowML/tree/main}.

\appendices

\section{Repositories Extracted for KG}~\label{app:source_acq}

To systematically investigate the attack mutation space, we conducted a comprehensive search of GitHub repositories containing relevant implementations. We began by constructing a base set of keywords for each attack, then expanded these keywords with variations to capture different naming conventions and implementation styles. Each combination was used as a query to search GitHub, and the resulting repositories were collected into a candidate set. This structured process ensured broad coverage of potential attack implementations and minimized the risk of overlooking relevant variants. A list of searched keywords are given in~\autoref{tab:search}.

\begin{table*}[t!]
\centering
\caption{Attack keywords used in search and number of repositories found}
\label{tab:search}
\scalebox{1}{% adjust 0.9 → smaller or larger as needed
\begin{tabular}{@{}p{2.5cm} p{3.5cm} p{5.3cm} r@{}}
\toprule
\textbf{Attack Name} & \textbf{Base Keywords} & \textbf{Variations} & \textbf{Repos} \\
\midrule
TCP DoS & TCP & 
Flood, DoS, Denial of Service, Attack & 2333 \\
\midrule
HTTP DoS & 
HTTP, HTTP GET, HTTP POST, HTTP Request & 
Flood, DoS, Denial of Service, Attack, Bombardment, Overload & 2935 \\
\midrule
Brute Force & 
SSH Dictionary, SSH Brute Force, SSH Brute-Force, SSH Password Guessing, SSH Password Cracking, SSH, SSH Login Guessing, SSH Password Spraying & 
Attack, Technique, Method, Tool, Hack, Crack, Attempt, Threat, Vulnerability, Exploit, Breach, Brute Force, Dictionary Attack & 2585 \\
\bottomrule
\end{tabular}
}
\end{table*}

\section{Evaluating LLM Effectiveness in Strategy Extraction}\label{app:llm_evl}
We evaluate \projectTitle’s ability to enumerate and extract attack strategies through a Named Entity Recognition (NER) task, as introduced in \autoref{subsec:kg_construction}, since accurate identification of strategy entities from heterogeneous repository documentation is essential for constructing the attack strategy \ac{KG}. From a corpus of 7,853 open source implementations across three attack families, we sample 150 repositories for detailed evaluation, balancing annotation cost and diversity. The sample covers four representative scenarios observed in practice: structured documentation, unstructured documentation, empty inputs to test robustness against hallucinations~\cite{huang2025survey}, and irrelevant content from mistakenly retrieved auxiliary repositories (for example, an HTTP Redis pooler returned for an HTTP DoS query~\cite{serverless-redis-http}).

We compare three models, GPT 3.5 and GPT 4o, which are general-purpose \acp{LLM} shown to be effective in security-related extraction tasks~\cite{hu2024llm}, and Gollie~\cite{sainz2023gollie}, a task-specific NER model. Following standard practice~\cite{sainz2023gollie, bogdanov2024nuner, garcia2024gpt}, we report precision, recall and F1 score (\autoref{tab:model_results}), where Recall measures coverage of human-annotated strategy entities and Precision measures agreement with those annotations. GPT 4o attains the highest Recall at 90.91\%, indicating strong coverage of relevant strategies across diverse documentation formats. 
Precision may appear lower, at 53.16\%; however, this estimate should be interpreted as a conservative lower bound rather than as evidence of a fundamental limitation.  We adopt ``exact string matching'' for evaluation, meaning semantically equivalent extractions are counted as false positives (e.g., ``Port Randomisation'' vs. ``Random Port Selection''). We deliberately report this worst-case metric, as (to the best of our knowledge) no established measure of semantic closeness exists that avoids introducing subjective bias. In practice, semantically-similar strategies are merged via embedding-based clustering (\autoref{subsec:kg_compression}), substantially reducing redundancy with minimal manual effort. Hallucinations are further mitigated through fixed seeds and structured output schemas (\autoref{subsec:repo_contruction}). In our setting, high Recall is crucial, because missed strategies create permanent gaps in the \ac{KG} and limit downstream generalization, whereas superfluous-but-related entities can be pruned at low manual cost, yielding higher effective Precision after clustering.

\begin{table}[t!]
\centering
\caption{Performance results for the Strategy-NER task.}
\label{tab:model_results}
\begin{tabular}{lccc}
\toprule
\textbf{Model} & \textbf{Precision} & \textbf{Recall} & \textbf{F1-Score} \\
\midrule
GPT-4o mini   & 0.5316 & \cellcolor{green!30}\textbf{0.9091} & 0.6709 \\
GPT-3.5 Turbo & 0.5981 & 0.5541 &  0.5753 \\
Gollie 13B    & 0.5417 & 0.0563 & 0.1020 \\
\bottomrule
\end{tabular}
\end{table}

\section{Feature Extraction Algorithm}\label{app:fe_extract}
For completeness, we provide the two core algorithms used in \projectTitle's feature extraction procedure mentioned in~\autoref{subsec:kg_augmented}. Algorithm~\ref{algo:update} specifies the update rule for a single feature statistic. Given a new observation $x_i$, it recursively updates the cumulative sum ($CS$), sample count ($W$), running mean ($r$), sum of squared residuals ($SSR$), and standard deviation ($\sigma$), without storing the entire history of observations. This formulation combines Welford’s algorithm~\cite{welford1962note} with the approach in Kitsune~\cite{mirsky2018kitsune}, enabling constant-time updates with minimal memory requirements. The complete update algorithm is given in~\autoref{algo:update_all}.

\begin{algorithm}[t!]
\caption{Register Network Statistics at Time $t$}
\begin{algorithmic}[1]
\Function{Update}{$CS_{t-x}, W_{t-x}, SSR_{t-x}, \sigma_{t-x}, x_i$}
    \State $CS_t \gets CS_{t-x} + x_i$ 
    \State $W_t \gets W_{t-x} + 1$ 
    \State $r_t \gets \frac{CS_t}{W_t}$ 
    \State $SSR_t \gets SSR_{t-x} + (x_i - r_t)^2$ 
    \State $\sigma_t \gets \sqrt{\frac{SSR_t}{W_t}}$ 
    \State \Return $CS_t, W_t, r_t, SSR_t, \sigma_t$
\EndFunction
\end{algorithmic}
\label{algo:update}
\end{algorithm}

\begin{algorithm}[t!]
\caption{Online Algorithm to Extract Relevant Feature Values for Packet at Time $t$}
\begin{algorithmic}[1]
\Function{ExtractFeatures}{packet $p$}
    \State $H_1, H_2 \gets p.identifiers$ \Comment{$H_1$: source identifier, $H_2$: destination identifier}
    \State $V_{H_1,H_2} \gets V_{channel}(H_1, H_2)$ 
    \State $V_{H_2} \gets V_{destination}(H_2)$ 
    \State $\Delta V \gets [ \ ]$ 
    
    \ForAll{$v$ in $V_{H_1,H_2}$}
        \State $x_i \gets p.f_i$ \Comment{Get corresponding value to update}
        \State $W_{t-x}, CS_{t-x}, r_{t-x}, SSR_{t-x}, \sigma_{t-x} \gets v$ 
        \State $\Delta CS \gets CS_{t-x}$ \Comment{Store value at previous timestep}
        \State $W_t, CS_t, r_t, SSR_t, \sigma_t \gets$ \Call{Update}{$CS_{t-x}, W_{t-x}, SSR_{t-x}, \sigma_{t-x}, x_i$}
        \State $v \gets W_t, CS_t, r_t, SSR_t, \sigma_t$ 
        \State $\Delta CS \gets |CS_t - \Delta CS|$ 
        \State $\Delta V \gets \Delta V \cup \{\Delta CS\}$ 
    \EndFor
    
    \ForAll{$(i, v)$ in \Call{enumerate}{$V_{H_2}$}}
        \State $x_i \gets \Delta V[i]$ \Comment{Get change in value}
        \State $W_{t-x}, CS_{t-x}, r_{t-x}, SSR_{t-x}, \sigma_{t-x} \gets v$ 
        \State $W_t, CS_t, r_t, SSR_t, \sigma_t \gets$ \Call{Update}{$CS_{t-x}, W_{t-x}, SSR_{t-x}, \sigma_{t-x}, x_i$}
        \State $v \gets W_t, CS_t, r_t, SSR_t, \sigma_t$ \Comment{Store updated statistics}
    \EndFor
    
    \State \Return $V_{H_1,H_2}\| V_{H_2}$ \Comment{Return updated feature values}
\EndFunction
\end{algorithmic}
\label{algo:update_all}
\end{algorithm}

Building on this primitive,~\autoref{algo:update_all} describes the end-to-end packet-level workflow for online feature extraction. For each packet, it retrieves the relevant feature spaces associated with the source–destination channel and the destination host, updates their statistics by repeatedly invoking~\autoref{algo:update}, and computes incremental values to capture short-term dynamics. The updated feature vectors are then returned for subsequent analysis. In summary,~\autoref{algo:update} provides the atomic update mechanism for individual features, while~\autoref{algo:update_all} integrates this mechanism into the complete packet-processing pipeline, enabling efficient and scalable extraction of descriptive statistics from network traffic in real time.

\section{Overview of Extracted Features}\label{app:feature_summary}
\autoref{tab:features} summarizes the extracted feature space used in our experiments. The features were designed according to the rules specified in \projectTitle, covering a broad range of traffic characteristics including flow-level metrics, packet size statistics, temporal dynamics, and protocol-specific behaviors. In addition to low-level features (e.g., TCP flags, inter-arrival times, retransmissions), we also incorporate higher-level protocol semantics (e.g., HTTP request methods, SSH authentication counts) to capture application-layer behaviors.

\begin{table*}[t!]
\centering
\caption{\textbf{\KGFE\ Feature Space of \projectTitle}. Features for HTTP, TCP DoS and Brute Force. ICMP features are included due to reconnaissance strategies found in examined repositories. The feature space is organized according to their underlying characteristics.}
\label{tab:features}
\small
\begin{tabular}{@{}%
    p{3.3cm}%
    p{4.8cm}%
    c%
    p{6.5cm}@{}}
\toprule
\textbf{Category} & \textbf{Statistics (Examples)} & \textbf{\# Features} & \textbf{Description} \\
\midrule
Connection Metrics 
    & Duration, keep-alive count, idle connections 
    & 8 
    & Connection lifespan and persistence patterns \\
Packet Volume 
    & Total count, size, inbound/outbound rates 
    & 18 
    & Directional packet counts and transmission rates \\
Inter-Arrival Time 
    & Total time, minimum, maximum, variation 
    & 8 
    & Temporal spacing between packets \\
Throughput 
    & Transmission rate, throughput 
    & 4 
    & Data transfer rates (bytes/sec) \\
TCP Flags 
    & SYN, ACK, RST, FIN, PSH, URG, XMAS, NULL 
    & 16 
    & Control flags indicating connection state and anomalies \\
TCP State 
    & State distribution (0--6), current state 
    & 8 
    & TCP state machine progression \\
TCP Flow Values 
    & Window size, header values
    & 28
    & Receiver buffer management and congestion signals \\
TCP Quality 
    & Retransmissions, packet loss, checksums, fragments 
    & 12 
    & Transmission reliability and error indicators \\
Time-to-Live (TTL) 
    & TTL (average, spread, min, max), packet count 
    & 14 
    & IP hop count distribution (routing path) \\
HTTP Methods 
    & GET, POST, PUT, DELETE, HEAD, PATCH, OPTIONS, TRACE 
    & 16 
    & Request type distribution \\
HTTP Status 
    & Success (2xx), client error (4xx), server error (5xx) 
    & 6 
    & Response code categories \\
HTTP Statistics 
    & Payload size, header size (average, spread, count) 
    & 18 
    & HTTP message size statistics \\
HTTP Protocol Anomalies 
    & Malformed headers, HTTP/2 frames (RST\_STREAM, GOAWAY) 
    & 22 
    & Protocol violations and HTTP/2-specific events \\
HTTP Timing 
    & Connection rate, authentication attempts 
    & 4 
    & Application-layer session dynamics \\
SSH Protocol 
    & Packet size/count (average, spread), key exchange attempts
    & 20 
    & SSH session establishment and authentication patterns \\
ICMP 
    & Packet count, error messages, ping requests 
    & 6 
    & Network diagnostics and reconnaissance indicators \\
Request/Response Timing 
    & Request interval, response interval 
    & 4 
    & Application-layer interaction timing \\
Connection Dynamics 
    & Establishment/termination attempts per second 
    & 4 
    & Connection lifecycle event rates \\
\bottomrule
\multicolumn{4}{@{}p{\textwidth}@{}}{\footnotesize
\textit{Total: 214 features (108 channel-level + 106 destination-level).}} \\
\end{tabular}
\end{table*}

\section{Hyperparameter Selection}\label{app:hyperparam}
We performed grid search over the parameters listed in \autoref{tab:grid_search_params}. Each model was tuned to maximize Recall under the constraint $\text{FPR} \leq 0.1\%$, selecting the lowest \ac{FPR} that achieved at least 90\% Recall. This avoids the ``benchmark lottery”~\cite{dehghani2021benchmark}, ensuring that performance reflects the method rather than arbitrary hyperparameters. For GMM, we varied $n\_components \in {3,5,7,10,20,30,40}$ and covariance type $\in {\ttype{diag}, \ttype{spherical}}$ (14 combinations). For DA, we searched over hidden ratios ${0.1,0.3,0.5}$, learning rates ${0.0001,0.001}$, corruption levels ${0.05,0.1,0.2}$, activations ${\ttype{relu},\ttype{tanh}}$, and $l_2$ regularization ${0.0001,0.001}$ (72 combinations). For \KITML, we tuned maxAE ${1,10}$, learning rate ${0.001,0.01}$, and hidden ratio ${0.25,0.5}$ (8 combinations). This was repeated 2$\times$ for different datasets.

\begin{table*}[h!]\small
\centering
\caption{Grid search parameters and number of combinations for each model, tested at $\text{FPR}_{te}$=0, 0.01, 0.1\%.}
\label{tab:grid_search_params}
\begin{tabular}{|l|p{8cm}|c|}
\hline
\textbf{Model} & \textbf{Parameters} & \textbf{Combinations} \\
\hline
GMM & 
\ttype{n\_components}: [3, 5, 7, 10, 20, 30, 40] \newline
\ttype{covariance\_type}: [diag, spherical]
& 14 \\
\hline
DA & 
\ttype{hidden\_ratio}: [0.1, 0.3, 0.5] \newline
\ttype{learning\_rate}: [0.0001, 0.001] \newline
\ttype{corruption\_level}: [0.05, 0.1, 0.2] \newline
\ttype{activation}: [relu, tanh] \newline
\ttype{l2\_reg}: [0.0001, 0.001]
& 72 \\
\hline
\KITML & 
\ttype{maxAE}: [1, 10] \newline
\ttype{learning\_rate}: [0.001, 0.01] \newline
\ttype{hidden\_ratio}: [0.25, 0.5]
& 8 \\
\hline
\end{tabular}
\end{table*}

\section{Additional Dataset Details}\label{app:dataset details}
In this paper, we evaluate attack variants organized by the categories defined in~\autoref{sec:issues}. The breakdown of evaluated and collected variants is presented in~\autoref{table:variants}. Note that some variants overlap and do not fall into a single category. In the remainder of this section, we describe the collection and simulation of the variants used in this study.

\begin{table}[t]\small
\centering
\caption{Summary of Attack Variants used for evaluation.}
\label{table:variants}
\setlength{\tabcolsep}{3pt}
\begin{tabular}{cll}
\textbf{Category} & \textbf{Variant Name} & \textbf{Dataset} \\
\toprule
\multirow{5}{*}{\rotatebox{90}{Out-of-Dimension}} 
    & HTTP Overflow              & CAP       \\
    & HTTP Mal                   & CAP       \\
    & Nkiller2                   & CAP       \\
    & Brute Force  & IoT-23    \\
    & Brute Force           & CIDS-17    \\
    & Brute Force (P=0)      & Con   \\
    & DoS Golden Eye            & CIDS-17    \\
    & DoS Hulk                  & CIDS-17    \\
\midrule
\multirow{8}{*}{\rotatebox{90}{Non-Throughput}} 
    & DoS-Slowloris             & CIDS-17    \\
    & DDoS-Slowloris            & CIoT-23    \\
    & DoS-Slowhttptest           & CIDS-17    \\
    & Fiberfox AVB               & CAP       \\
    & Fiberfox BYPASS            & CAP       \\
    & Low-rate TCP               & CAP       \\
    & Brute Force (P=1)      & Con     \\
    & Fiberfox GET         & CAP       \\
\midrule
\multirow{16}{*}{\rotatebox{90}{Composite}} 
    & DoS-HTTP             & CIoT-23    \\
    & DDoS-HTTP           & CIoT-23    \\
    & DoS-SYN                    & CIoT-23    \\
    & DDoS-SYN                  & CIoT-23    \\
    & DoS-TCP                   & CIoT-23    \\
    & DDoS-TCP                   & CIoT-23    \\
    & DDoS-PSHACK               & CIoT-23    \\
    & DDoS-RSTFIN               & CIoT-23    \\
    & DDoS-SynIP         & CIoT-23    \\
    & DDoS-ACK Frag     & CIoT-23    \\
    & Fiberfox SLOW        & CAP       \\
    & Fiberfox STRESS      & CAP       \\

\bottomrule
\end{tabular}

\end{table}

\subsection{Simulated Attacks} \label{appx:sim_attacks}
We simulated 4 attack scenarios, applied to benign traffic from both the \ac{CIoT-23} and \ac{CIDS-17} datasets, including the aforementioned Nkiller2 attack. Each attack represents a real-world \textit{known} vulnerability. The simulations were conducted by taking an hour of benign traffic. For each simulation  we forced at least a 2x increase in rate compared to the original benign traffic (if not stated otherwise). The goal is to show that even with an increase in the average byte rate, those attacks still cannot be captured due to the insufficient discriminative power of the feature space. The configuration and setup details are given as follows: 
\begin{enumerate}
        \item \textbf{Low-rate TCP}:  Throttling the TCP SYN attack by sending packets in bursts of 1 second and an idle time of 1 second. Simulating a well-established Low-Rate TCP Targeted attack \cite{kuzmanovic2003low} targeting the TCP Retransmission Timeout (RTO) mechanism.
        \item \textbf{Nkiller2}: Setting TCP window size of TCP packets to 0 and sending packets at 2.2x rate as the benign traffic. Simulating the vulnerability \cite{CVE202432984, CVE20084609, CVE20091926}.
        \item \textbf{HTTP DoS (HTTP Overflow)}: Setting HTTP Accept-Language to overflow a long valid string, causing the server to spend excessive time processing (as described in~\cite{CVE-2024-39316}). The packets are sent at 2.2x rate as benign traffic.
        
        \item \textbf{HTTP DoS (Mal Header)}: Setting HTTP Accept-Encoding to manifold of invalid values which can cause the kernel to crash, simulating vulnerability \cite{CVE-2021-31166,CVE-2025-31650}. The packets are sent at 2.2x rate as benign traffic.

        % \item \textbf{HTTP Request Smuggling (Bypass Char 1)}: Injecting malicious commands into HTTP packet headers using bypass space technique (for details refer to \cite{PayloadsAllTheThings-CommandInjection}).
        % \item \textbf{HTTP Request Smuggling n (Bypass Space 2)}: Injecting malicious commands into HTTP packet headers using bypass space technique (for details refer to \cite{PayloadsAllTheThings-CommandInjection}).     
\end{enumerate}

\subsection{Captured Attacks}\label{appendix:captured}
The CAP dataset was collected in a controlled but realistic setting using two heterogeneous hosts: a MacBook running Sonoma 14.1 and a Dell PC running Ubuntu 14. The Dell PC was configured with an \ttype{nginx} HTTP service when HTTP DoS attacks were executed. The Dell machine acted as the victim, and all traffic was captured directly on the victim using Wireshark to minimize measurement artifacts and ensure that the traces reflect the actual network load experienced by the service. The MacBook acted as the attacker and generated malicious traffic by replaying five different HTTP DoS attack strategies using Fiberfox~\cite{fiberfox}.

To avoid dataset bias and ensure realism, we (i) used commodity hardware and standard operating systems instead of emulators, (ii) deployed widely used services (nginx for HTTP) with default configurations, and (iii) allowed background system processes and benign traffic to coexist with attacks during collection. This setup prevents artificial isolation of attack traffic and preserves natural protocol dynamics such as connection establishment, response codes, and timing behavior. In addition, we avoided synthetic traffic generation tools beyond the attack scripts themselves, thereby reducing the risk of artifacts in packet structure or timing distributions. Each attack strategy was executed for 20 minutes (as specified in the \ttype{README.md} of Fiberfox~\cite{fiberfox}), resulting in five distinct attack scenarios:
\begin{enumerate}
\item \ttype{--AVB}: Issues HTTP GET packets into an open connection with long delays between send operations.
\item \ttype{--GET}: Sends randomly generated HTTP GET requests over an open TCP connection.
\item \ttype{--STRESS}: Sends a sequence of HTTP requests with a large body over a single open TCP connection.
\item \ttype{--BYPASS}: Sends HTTP GET requests over an open TCP connection and reads responses back.
\item \ttype{--SLOW}: Similar to STRESS, but issues HTTP requests while keeping the connection active by reading back a single byte and sending additional payload with timed delays.
\end{enumerate}

\paragraph{Con} This dataset was provided by external collaborators (not produced as part of this work) and was collected using SSH Patator~\cite{lanjelot_patator} to evaluate generalization against SSH variants.
\ttype{--SSH\_PERSISTENT}: If set to 0, it creates a new connection for each attempt; otherwise, it sends multiple brute force attempts through the same connection until termination.

\section{Patching Kitsune}~\label{app:kitsune_patch}

When implementing baselines, we referred to the Kitsune open-source implementation at~\cite{KitsunePy}. We observed an exponential increase in runtime during the feature extraction process in Kitsune. Our analysis revealed that the covariance update underwent an unintended double decay, which deviates from the algorithm described in the original paper~\cite{mirsky2018kitsune}, leading to the runtime growth. To ensure fair comparison, we corrected this issue, and all experiments in this work were conducted using the patched version of Kitsune. \autoref{tab:kitsune_comparison} presents performance results on the original Mirai dataset. Our results are comparable to those reported by Arp et al.\cite{arp2022and}, but differ from those in the original Kitsune paper, as a number of features had been removed by the authors~\cite{KitsunePyIssue12}.

\begin{table}[t]%\small
    \centering
    \caption{Comparison of Kitsune Before and After Fix at FPR=0 on the Kitsune Mirai dataset}
    \scalebox{0.85}{
    \begin{tabular}{lrr}
        \toprule
        Metric & Kitsune Before Fix & Kitsune After Fix \\
        \midrule
        True Positives (TP)  & 560,735  & 565,390  \\
        True Negatives (TN)  & 66,620   & 66,620   \\
        False Positives (FP) & 0        & 0        \\
        False Negatives (FN) & 81,781   & 77,126   \\
        Precision            & 1.0000   & 1.0000   \\
        Recall               & 0.8727   & \cellcolor{green!20} 0.8791 \textbf{↑ (+0.0064)} \\
        F1-score             & 0.9320   & \cellcolor{green!20} 0.9361 \textbf{↑ (+0.0041)} \\
        \bottomrule
    \end{tabular}
    }
    \label{tab:kitsune_comparison}
\end{table}

\end{document}